\documentclass[traditabstract]{aa}
\usepackage{latexsym,graphicx,natbib,amssymb,array,aas_macros,txfonts}
\usepackage{txfonts}

\newcommand{\msun}{M_{\odot}}
\newcommand{\mecl}{M_{\rm ecl}}
\newcommand{\rh}{r_{\rm h}}
\newcommand{\rhoecl}{\rho_{\rm ecl}}
\newcommand{\fb}{f_{\rm bin}}
\newcommand{\nb}{N_b}
\newcommand{\ncms}{N_{\rm cms}}
\newcommand{\nbin}{n_{\rm bin}}
\newcommand{\mpc}{M_{\odot}\;{\rm pc}^{-3}}
\newcommand{\mbar}{\overline{m}}
\newcommand{\Odyn}{\Omega_{\rm dyn}^{\mecl,\rh}}
\newcommand{\Dfin}{\Phi_{\log_{10}a}^{\mecl,\rh}(t)}
\newcommand{\Dobs}{\Phi_{\log_{10}a}^{\rm obs}}
\newcommand{\tcri}{t_{\rm cr,initial}}
\newcommand{\chimin}{\chi^2_{\rm min}}
\newcommand{\probchi}{P(\geq\chi^2_{\rm min})}

\begin{document}

\title{Inverse dynamical population synthesis}
\subtitle{Constraining the initial conditions of young stellar clusters by studying their binary populations}
\titlerunning{Inverse Dynamical Population Synthesis}
      
\author
{
  Michael Marks\inst{\ref{inst1},\ref{inst2},}\thanks{Member of the International Max Planck Research School (IMPRS) for Astronomy and Astrophysics at the Universities of Bonn and Cologne; e-mail: mmarks@astro.uni-bonn.de (MM)} \and Pavel Kroupa\inst{\ref{inst1}}
}

\institute{Argelander Institute for Astronomy, University of Bonn, Auf dem H\"ugel 71, 53121 Bonn, Germany\label{inst1} \and  Max-Planck-Institut f\"ur Radioastronomie, Auf dem H\"ugel 69, D-53121 Bonn, Germany\label{inst2}}

\authorrunning{Michael Marks \& Pavel Kroupa}

\date{Received <date> / Accepted <date>}

\abstract{
Binary populations in young star clusters show multiplicity fractions both lower and up to twice as high as those observed in the Galactic field. We follow the evolution of a population of binary stars in dense and loose star clusters starting with an invariant initial binary population and a formal multiplicity fraction of unity, and demonstrate that these models can explain the observed binary properties in Taurus, $\rho$~Ophiuchus, Chamaeleon, Orion, IC 348, Upper Scorpius A, Praesepe, and the Pleiades. The model needs to consider solely different birth densities for these regions. The evolved theoretical orbital-parameter distributions are highly probable parent distributions for the observed ones. We constrain the birth conditions (stellar mass, $\mecl$, and half-mass radius, $\rh$) for the derived progenitors of the star clusters and the overall present-day binary fractions allowed by the present model. The results compare very well with properties of molecular cloud clumps on the verge of star formation. Combining these with previously and independently obtained constraints on the birth densities of globular clusters, we identify a weak stellar mass -- half-mass radius correlation for cluster-forming cloud clumps, $\rh/{\rm pc}\propto(\mecl/\msun)^{0.13\pm0.04}$. The ability of the model to reproduce the binary properties in all the investigated young objects, covering present-day densities from $1-10$~stars~pc$^{-3}$ (Taurus) to $2\times10^4$~stars~pc$^{-3}$ (Orion), suggests that environment-dependent dynamical evolution plays an important role in shaping the present-day properties of binary populations in star clusters, and that the initial binary properties may not vary dramatically between different environments.
}

\keywords{
binaries: general - star clusters: general - globular clusters: general - open clusters and associations: general - methods:numerical
}
\maketitle

\section{Introduction}
\label{sec:intro}
It has been observationally established that the properties of binary populations in star clusters depend on environment: the frequency of binaries is lower in denser and older populations. For instance, the Taurus star-formation region, which is about $1$~Myr old, has a very low density ($1-10$~stars~pc$^{-3}$) and a binary population that is about twice the size \citep{Ghez1993,Leinert1993,Simon1995,Koehler1998,Kraus2009} of those for solar-type main sequence stars in the Galactic field \citep[the canonical sample used here for comparison,][]{DuqMay1991}. On the other hand, the similarly aged Orion Nebula Cluster has a much higher density ($2\times10^4$~stars~pc$^{-3}$) and its binary population is smaller than that in the field \citep{Reipurth2007}. The density of IC 348 is comparable to that of NGC 2024, but the former is $3-5$~times older. IC~348 has a smaller binary fraction than NGC 2024 \citep{Duchene1999a,Levine2000}.

Similar behaviour is reported for older open \citep{Sollima2010} and globular clusters \citep{Sollima2007} of the Galaxy, inferred from anti-correlations between the binary proportion and cluster age, as well as between binary proportion and cluster luminosity (i.e. mass).

Theoretical approaches to explaining these differences can be categorized into two major groups. One proposes that the \emph{formation} of binaries dependends on the environment. In this case the prevalent conditions in the star-forming cloud can determine the outcome of binary star formation \citep{DurisenSterzik1994,Sterzik2003}. The second group of scenarios propose environment-dependent \emph{dynamical evolution}, in which the properties of binary populations seen in star clusters cannot be assumed to be primordial, i.e. they are not directly related to the birth properties of binaries. In these scenarios, the characteristics are modified by interactions between the stars and binaries. Even if the clusters are still very young ($\lesssim1$~Myr), the binary properties can strongly deviate from the birth conditions produced in the star-formation process since the binary proportion is reduced rapidly and orbital parameters are altered efficiently over only a few initial crossing-times \citep{Kroupa1995a,Duchene1999a,Fregeau2009,Parker2009}.

In this paper, we investigate whether dynamical evolution is able to explain observations of the multiplicity properties in very different environments, starting with an assumed universal initial binary population in star clusters \citep{Kroupa1995b}. In particular, we apply the concept of inverse dynamical population synthesis to constrain the properties of star clusters at their birth.

In Sect.~\ref{sec:binaries}, we present the model and the procedure to analyse the eight investigated regions we describe in Sect.~\ref{sec:sample}. In Sect.~\ref{sec:results}, we present the outcome of our analysis and Sect.~\ref{sec:sum} summarizes and discusses our results.

\section{Binaries in star clusters}
\label{sec:binaries}
\subsection{Stimulated evolution, dynamical equivalence, and the age-density degeneracy}
\label{sec:stim}
A stellar binary population in a star cluster evolves as binaries interact gravitationally with other stars or binaries. As energy and momentum between these centre-of-mass systems is exchanged, binary orbital parameters are altered or, if the transferred amount of energy is sufficient, a binary dissolves. Generally during encounters, a \emph{soft} binary, which has an internal binding energy that is lower than the kinetic energy of a typical centre-of-mass system in the cluster, becomes on average less bound (softens), while a \emph{hard} binary with a higher binding energy becomes more strongly bound \citep[hardens,][]{Heggie1975,Hills1975}. This \emph{stimulated evolution} of binaries depletes the binary population. The process is most efficient in wide low-energy systems, which are those most likely to be disrupted.

\begin{figure}
 \resizebox{\hsize}{!}{
   \includegraphics[width=0.45\textwidth]{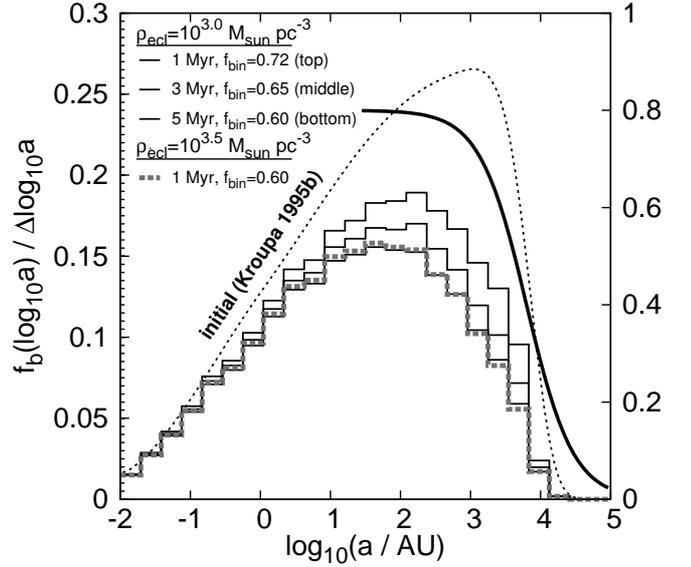}
 }
 \caption{The $\Omega$-operator in action (eq.~\ref{eq:odyn}, thick solid S-shaped curve, schematic, right ordinate). Its exact shape depends on the cluster volume-density within the half-mass radius, $\rhoecl$, and time, $t$, for stimulated evolution. Multiplying the respective operator by the initial binary population \citep[thin dotted line]{Kroupa1995b} for a cluster with $\rhoecl=10^3\mpc$, and age, $t=1,3,$ and $5$~Myr, respectively, results in the three semi-major axis distributions (solid histograms from top to bottom). Most of the evolution has already taken place in the first Myr, where the binary fraction decreases from $100$ to $72$~percent in this particular cluster. The grey thick dashed line on top of the lowest black solid line is the resulting semi-major axis distribution for a $\rhoecl=10^{3.5}\mpc$ cluster after $1$~Myr of stimulated evolution. This shows that two clusters with densities $\rhoecl(1)<\rhoecl(2)$ can evolve into the same resulting distribution when we assume $t(2)<t(1)$ is appropriately chosen (owing to the age-density degeneracy). That is, the $\rhoecl(1)$ cluster at age $t(1)$ is dynamically equivalent to the $\rhoecl(2)$ cluster at age $t(2)$.}
 \label{fig:odyn}
\end{figure}

The resultant change in the orbital parameters for populations of binaries evolving in star clusters driven by stimulated evolution is quantified by \citet{MarksKroupaOh2011} by means of a comprehensive set of $N$-body computations of binary-rich star clusters. In these models, all stars are initially members of a binary. The component masses of late-type stars are selected randomly from the canonical stellar initial mass-function \citep[IMF,][]{Kroupa2001}. The IMF is conveniently described as a two-part power-law in the stellar regime,
\begin{equation}
 \label{eq:imf}
 \xi(m)\propto m^{-\alpha_i}\;{\rm where}\;\left\{
 \begin{array}{lr}
  \alpha_1=1.3 & 0.08\leq m/\msun\leq0.5,\\
  \alpha_2=2.3 & 0.5<m/\msun\leq150,
 \end{array}
 \right.
\end{equation}
with an average stellar mass $\mbar\approx0.4$ where $dN=\xi(m)dm$ is the number of stars in the mass interval $[m,m+dm]$. Random pairing results in late-type binaries having a flat initial mass-ratio distribution which is affected only in a minor way through pre-main sequence eigenevolution (fig.~17 in \citet{Kroupa2008} and fig.~2 in \citet{MarksKroupaOh2011}). The mass ratio distribution of late-type binaries remains approximately flat after dynamical processing. The orbital periods of the binaries are selected from an initial period distribution function that increases with increasing period and flattens at the longest periods \citep[Fig.~\ref{fig:odyn}]{Kroupa1995b}. This distribution agrees\footnote{Note that this initial distribution is \emph{not} a fit to any observed data, but was identified by inverse dynamical population synthesis (Sect.~\ref{sec:idps}). It is however based on an initial guess of a flat distribution that matches the pre-main sequence constraints.} with pre-main sequence and proto-stellar multiplicity data \citep{KroupaPetr2011,MarksKroupaOh2011}. The period distribution is assumed to be the universal outcome of binary formation \citep{Kroupa2011}.

Allowing this invariant initial binary population to evolve for $5$~Myr in clusters of different initial masses and sizes, \citet{MarksKroupaOh2011} find that for their set of models (of stellar mass in the computed clusters $\mecl\leq10^{3.5}\msun$ and half-mass radii $\rh\leq0.8$~pc), the evolution is driven by the initial cluster density, $\rhoecl=3\mecl/8\pi\rh^3$, i.e. models of the same $\rhoecl$ evolve their population in the same way. They are \emph{dynamically equivalent} (over the $5$~Myr integration time).

\citet{MarksKroupaOh2011} quantify a stellar dynamical operator, $\Odyn(t)$, which transforms an initial orbital-parameter distribution, $\Phi_{x,\rm in}$ (e.g., $x=$semi-major axis or $x=$mass-ratio), into an evolved distribution, $\Phi_{x}^{\mecl,\rh}(t)$, after some time, $t$, of stimulated evolution
\begin{equation}
 \Phi_{x}^{\mecl,\rh}(t)=\Odyn(t)\times\Phi_{x,\rm in}\;.
 \label{eq:odyn}
\end{equation}
Eq.~(\ref{eq:odyn}) allows the efficient calculation of the evolution of a binary population in a star cluster of a given initial density without the need for expensive individual $N$-body computations. Fig.~\ref{fig:odyn} depicts how the operator works. The evolution of the binary population has largely ended by $5$~Myr, the time for which the models in \citet{MarksKroupaOh2011} are computed. We note that there is an \emph{age-density degeneracy}, i.e. a high density cluster of age $t(1)$ may have the same binary population as a somewhat lower-density cluster of age $t(2)>t(1)$ since stimulated evolution in the latter cluster is less efficient.

In this paper, the considered orbital parameter distribution is always the distribution of semi-major axes, $\log_{10}a$, of the considered region. For a range of primary masses around $m_1$, it is defined as the distribution of binary-fractions $\fb=\nb/\ncms$ as a function of $\log_{10}a$
\begin{equation}
 \Phi_{\log_{10}a}^{\mecl,\rh}(m_1)=\frac{d\fb(\log_{10}a)}{d\log_{10}a}=\frac{1}{\ncms}\frac{d\nb(\log_{10}a)}{d\log_{10}a}\;,
\end{equation}
where $\nb$ and $\ncms$ are the number of binaries and centre-of-mass systems (singles+binaries), respectively. The total binary fraction equals the area below the distribution, i.e. the integral over the semi-major axis distribution is
\begin{equation}
 \fb(m_1)=\int\Phi_{\log_{10}a}^{\mecl,\rh}(m_1)\;d\log_{10}a\;.
 \label{eq:fb}
\end{equation}
For a description on the method and extraction of distribution functions, the reader is referred to \citet*{MarksKroupaOh2011}.

\subsection{Inverse dynamical population synthesis}
\label{sec:idps}
In this paper, the inverse operator, $(\Odyn)^{-1}(t)$, is applied to the observed binary distributions in clusters of age $t$ in order to see whether stimulated evolution in a cluster of initial density $\rhoecl$ can turn the universal initial distribution into a shape which serves as a parent function of the observational data.

This inverse dynamical population synthesis (IDPS) relies on assuming an invariant birth binary star population and matching an observed binary population with a dynamical model from which the initial conditions of the population in terms of its density are derived. \citet{Kroupa1995a} introduced this concept and demonstrated that physically relevant solutions are obtainable. He postulated that two stellar dynamical operators, $^P\Odyn$ and $^q\Odyn$, exist. The first one ($^P\Odyn$) evolves the period distribution function of a binary population through stellar-dynamical encounters, while the other one ($^q\Odyn$) evolves the mass-ratio distribution. As both distribution functions are not correlated for periods $>10^3$~days, solutions for $^P\Odyn$ and $^q\Odyn$ can only be obtained if these operators correspond to physical processes. \citet{Kroupa1995a} not only discovered the existence of these operators, but also demonstrated that they are equal, $^P\Odyn={^q\Odyn}$, i.e. the same process (stimulated evolution) shapes these distributions. Thus, a single operator, $\Odyn$, exists to turn both the initial period and the initial mass-ratio distribution into the observed shape simultaneously. It is given by a star cluster with $\mecl=128\msun$ and $\rh=0.8$~pc, which corresponds to the typical embedded cluster from which the majority of Galactic field binaries originate \citep[cf.~also][]{MarksKroupa2011}. \citet{Kroupa2002} and \citet{MarksKroupaOh2011} formalized the mathematical concept of $\Odyn$, which is here used for the first time to derive the initial conditions of observed star clusters.

\subsection{Modelling observed binary populations}
\label{sec:modelling}
We construct the observationally inferred semi-major axis distribution, $\Dobs$, from $N$ orbits distributed into $\nbin$ bins, and the evolved theoretical distribution, $\Dfin$, where $t$ is chosen to correspond to the age of the investigated region for which the $\rhoecl$ is known (eq.~\ref{eq:odyn}). The best-fit solution for $\rhoecl$ is then found by performing a $\chi^2$-test with $\nbin-1$ degrees of freedom, i.e. the sum of the squared deviations between $\Dobs$ and $\Dfin$ is minimized for $\rhoecl$
\begin{equation}
 \chi^2(\rhoecl) = \sum_{i=1}^{\nbin}\frac{\left(\Dfin-\Dobs\right)_i^2}{\left(\Delta\Dobs\right)_i^2} = {\rm min}\;,
\end{equation}
where $\Delta\Dobs$ is the error in the observational data in bin~$i$, which is assumed to be $\sqrt N$ errors for all data sets.

To estimate the likelohood that the resulting $\Dfin$ is indeed a parent function of $\Dobs$, we calculate the probability, $\probchi$, that a poorer fit occurs \citep{Press1994}. For instance, at a given $\chi^2_{\rm min}$, $P(\gtrsim\chi^2_{\rm min})=0.3$ means that if one rejects the null hypothesis (that ``$\Dfin$ is a parent function for $\Dobs$``), then there is a $30$~percent chance that $\Dfin$ is a parent function, even though it has been withdrawn. Confidence intervals around the best $\rhoecl$ are defined by the condition $P(\gtrsim\chi^2_{\rm min})=10$~percent, outside which the null hypothesis is falsely withdrawn with $\leq10$~percent chance only.

The $\chi^2$-test allows the identification of a typical birth density $\rhoecl$ for the considered region and via eq.~(\ref{eq:fb}) the overall present-day binary fraction can be assessed using the best-fit theoretical distribution.

While the dynamical operator acts on a population of binaries of all primary masses, the observations mostly probe masses of a restricted primary-mass range. As both the binary properties and their distributions are mass dependent \citep{Kouwenhoven2009,MarksKroupa2011}, this effect is important to consider. The analytical model allows us to extract only those binaries in the theoretical population that cover the same range of stellar masses as the observations \citep[see sect.~2.4 in][]{MarksKroupa2011}. However the effect on the results is negligible as the observational uncertainties are much larger than the mass-dependent effects discussed here.

\section{The young cluster sample}
\label{sec:sample}
We now introduce the young cluster sample that we analyse with the model described in Sect.~\ref{sec:modelling}. For each cluster, we compile information about their present-day masses and sizes and their binary population. We emphasize that when discussing the dynamical processing it is indispensible to distinguish between the dynamically active sub-clumps or individual clusters that always have radii of about $0.5$~pc and the whole coeval stellar population that may extend over a few pc (see also the discussion in Sect.~\ref{sec:sum}).

We ascertain from the literature whether the initial stellar mass distribution is compatible with the canonical IMF, which is used as input into the $N$-body models of \citet{MarksKroupaOh2011}. The theoretical distributions to compare with the observational data are evolved for the age of the respective region. Since these ages are often uncertain or an age spread is present, we use mostly upper age-limits, as indicated in Figs.~\ref{fig:taurus_oph} to~\ref{fig:praesepe_pleiades}, to construct the theoretical distribution. Owing to the age-density degeneracy (Fig.~\ref{fig:odyn}), this will result in lower limits to the inferred initial densities. In addition, when the ages are chosen to differ slighty the resulting change in the densities is smaller than the given confidence intervals, which are rather large given the observational uncertainties. Therefore, the results are considered reasonable estimates of the initial densities.

\subsection{Taurus}
\label{sec:taurus}
The Taurus star-formation region has a distance of $140$~pc \citep{Kenyon1994,Wichmann1998}, which is nearby, and extends over a region of $\left(100^{\circ}\right)^2$ on the sky. The age of its population is $\approx1$~Myr. Six distinct clusters of stars in Taurus were detected by \citet{Gomez1993} to have projected radii of $0.5-1$~pc with $\approx15$ stars or binaries (with a mass of about $6\msun$ for $\mbar\approx0.4$) each and a median separation between the stellar centre-of-mass systems within each cluster of $0.3$~pc. The individual clusters are separated by a few pc, thus do not interact dynamically. Taurus contains more than $300$ known members and the stellar density in each of its clusters is at $1-10$~stars~pc$^{-3}$ very low \citep{Luhman2009}. The N-body models of \citet{Kroupa2003} show that Taurus Auriga like clusters are largely unevolved and in particular that the binary properties should be similar to those at birth.

Taurus was originally proposed to be deficient in low-mass stars \citep{Briceno2002}, but many brown dwarfs have since been discovered \citep{Luhman2004T,Guieu2006}. \citet{Kroupa2003b} and \citet{ThiesKroupa2008} show that there is no need to invoke a non-canonical IMF in Taurus.

Among the $43$~observed primary stars in Taurus, \citet{Ghez1993} find $16$~binaries with angular separations between $16-252$~AU. The binary fraction is thus $37\pm9$~percent over this separation range. \citet{Simon1995} find $22$ binaries and $4$ triples among $47$~systems. The binary frequency in the range $3-1400$~AU is at least $1.6\pm0.3$~times the value of the canonical Galactic field sample. \citet{Kraus2009} find $27$ wide binaries in Taurus in the separation range $500-5000$~AU, which is consistent with a log-flat distribution but inconsistent with the Galactic field log-normal distribution. Here the combined data of \citet{Leinert1993} and \citet{Koehler1998} are used, who identified in total $74$~binaries or multiples among $174$~systems. The binary fraction of $42.5\pm4.9$~percent is larger by a factor of $1.93\pm0.26$ than that of solar-type main sequence stars over the accessible separation range.

\subsection{$\rho$ Ophiuchus}
\label{sec:oph}
The $\rho$ Ophiuchi cloud is a complex of dark nebulae and molecular clouds with ongoing star formation located at about $\approx120-160$~pc distance from the Sun \citep[$140$~pc is adopted]{Chini1981,KnudeHog1998,Motte1998}. Its densest part is the L1688 dark cloud, which is itself structured into the six major clumps Oph-A to Oph-F, each containing a few tens of solar mass in young stellar objects (YSOs) and each having a diameter of $\approx0.3$~pc \citep{Stamatellos2007}. \citet{Loren1990} use a total mass (YSOs + gas) for each clump of $\approx200\msun$. The ages of the stars are inferred to lie in the range $0.3-1$~Myr \citep{Wilking1989,GreeneMeyer1995,LuhmanRieke1999}. \citet{Bontemps2001} find, after correcting for an assumed binary fraction of $75$~percent, a stellar mass-function for young stellar objects (YSOs) that is consistent with the slope of the canonical stellar IMF for low masses ($0.055-0.55\msun$) and agrees with earlier \citep{Comeron1993,Strom1995,Motte1998,Johnstone2000} and later \citep{Tachihara2002,Stanke2006} studies. A somewhat steeper mass-function than the canonical IMF is observed for higher stellar masses \citep[$\alpha_2=1.7$ in eq.~\ref{eq:imf},][]{Bontemps2001}. The same authors estimate that there is $92\msun$ of mass in YSOs in L1688 and given an average radius of $0.4$~pc for the whole region they find a stellar mass density of $340\mpc$. The extents of the clumps are about $0.1-0.2$~pc. The total gas mass of the region is between $550\msun$ \citep{WilkingLada1983} and $1500\msun$ \citep{Liseau1995}. \citet{Ratzka2005} state a value of $600\msun$ for the total mass within a region of $1\times2$~pc.

The YSO content of L1688 has been intensively surveyed in the past two decades. \citet{Ghez1993} observed a binary frequency for T~Tauri stars of $29\pm11$~percent (sample size of 21 stars only) in the linear separation range $16-252$~AU, which is significantly higher than that in the field. \citet{Simon1995} find $10$ binaries, $2$ triples, and $1$ quadruple among $35$ T~Tauri primaries in $\rho$~Oph in the separation range $3-1400$~AU, i.e. a binary fraction of $37\pm10$~percent. Since they do not perform a completeness correction, this is \emph{at least} $1.1$~times larger than that in the field over the same range. \citet{Duchene1999a} correct \citet{Ghez1993}'s and \citet{Simon1995}'s data for incompleteness and re-evaluated the \emph{companion-star fraction} of $\rho$~Oph to be $26\pm5$~percent. \citet{Barsony2003} find $24\pm11$~percent within $14-154$~AU among $80$~primaries in their restricted sample. \citet{Duchene2004} observe $41$~YSOs in a deep near-infrared survey, finding $14$ companions to $11$ primaries, i.e. a companion star fraction of $34\pm7$~percent.

In this work, the data by \citet{Ratzka2005} are used, having observed the to-date largest sample of YSOs in $\rho$~Oph for their multiplicity properties. They find 112 single, 43 binaries, and 3 triples among their $158$-object-sized sample, resulting in a binary fraction of $29.1\pm4.3$~percent in the separation range from $18$ to $900$~AU. This is slightly larger than the corresponding field value ($23.5\pm4.8$~percent). Their semi-major axis distribution is indistinguishable from those published in \citet{Ghez1993}, \citet{Simon1995}, and \citet[see Fig.~\ref{fig:taurus_oph}, lower left panel]{Barsony2003}.

\subsection{Chamaeleon}
\label{sec:cham}
The star-formation region Chamaeleon is $160-170$~pc away from the Sun \citep{Whittet1997,Wichmann1998,Bertout1999} and has a median age of about $2$~Myr \citep{Luhman2004}. Cha~I and Cha~II are the main cloud complexes in Chamaeleon where the majority of stars are being formed. The IMF is compatible with that in the field as well as IC~348 and the Orion Nebula Cluster \citep{Comeron2000,Luhman2008}. The total mass in the dark clouds of Chamaeleon is about $5000\msun$, Cha~I containing 237 known members, and 50 members having been identified in Cha~II\citep{Luhman2008}. The total stellar mass in Cha~I is about $100\msun$, most of it being contained within a radius of $1^{\circ}$ \citep{Luhman2007}, corresponding to $2.88$~pc at the distance of Chamaeleon.

Among $77$~observed primaries, \citet{Koehler2001} find after correction for background stars $10.3$ binaries and $0.5$ triples, respectively. The binary fraction in the separation range $10-480$~AU is thus $14$~percent, about nine percent smaller than the 23 percent determined for solar type stars in the Galactic field. Interestingly, this deficit occurs only in their two largest separation bins, while the smaller separation bins neatly agree with the observed field separation distribution. This deficit of wide binaries is also reported by \citet{Brandeker2006}. The most recent study of \citet{Lafreniere2008}, which finds $30$~binaries and $7$~triples among $126$~primaries, does not however suggest that there is a strong depletion for the widest binaries. These authors provide projected angular separations for their observed binaries and triples, but they do neither show nor discuss a separation distribution. \citet{Lafreniere2008}'s data is used here and the distribution agrees also at the largest separations with the field distribution.

\subsection{Orion Nebula Cluster}
\label{sec:orion}
The Orion Nebula Cluster (ONC) is a young cluster ($\lesssim1$~Myr) that is part of the Orion complex at a distance of about $414-343$~pc \citep{Menten2007,Kraus2007} from the Sun. With its high central density, it is probably the densest nearby star-formation region. \citet{Hillenbrand1998} review earlier work on the ONC and determine its main properties. They recover $3500$~centre-of-mass systems in total out to a radius of $2.5$~pc. They calculate that if the cluster were in virial equilibrium a total mass of $\approx4500\msun$ would be needed within $2$~pc, which differs from the observed stellar mass within this radius of $1800\msun$. In the core of the ONC (size $0.16-0.21$~pc), the stellar density is about $2\times10^4$~stars pc$^{-3}$. Stellar masses range from $0.1$ to $50\msun$ with an average mass of $0.8\msun$ with most massive stars being found in the centre. The core contains the Trapezium system consisting of (at least) $6$~stars with a total mass of $88.4\msun$ \citep{Hillenbrand1997}. Brown dwarfs have since been detected in the centre of the cluster \citep{Hillenbrand2000,Luhman2000,Lucas2001,Muench2002}, so that the mean mass may decreases to $0.5\msun$ \citep{Scally2005}, in good agreement with the canonical IMF.

Existing dynamical $N$-body models suggest that the ONC must have been far more compact initially to appear in its present shape. \citet{Kroupa2000} computed models that start in virial equilibrium and are either dynamically hot (expanding) or cold (collapsing), with the same realistic initial binary population used here and expulsion of residual gas. He searched which initial conditions reproduce the observed velocity dispersion and observed number of stars in the core simultaneously. The models initially in virial equilibrium were basically excluded, owing to the large number of stars needed to match the observational properties. Models starting with $3500$ to $10000$ stars initially (corresponding to $1400$ to $4000\msun$ for $\mbar=0.4\msun$) that collapse and expand allow for ONC progenitors with half-mass radii of $0.5-1.2$~pc and $<0.5$~pc, respectively. \citet{Scally2005} compute very similar models to the \citet{Kroupa2000} ones but without gas and binaries. Their computations agree in their main results, although virial equilibrium models most closely reproduce the observed density profile. In this case, the initial density may have been one to two orders of magnitude higher than that today. \citet{PflammAltenburg2006} investigate the decay of the ONC core. Models with initially $40$~massive stars ($>600\msun$), which is the expected number if stars in the ONC were selected from the canonical IMF at birth, in a region of $0.025$~pc only end up with about the observed number of massive objects (10 stars above $5\msun$) and display a Trapezium structure.

The binary population of the ONC has been found to be about half as large as that in Taurus \citep{Prosser1994,Padgett1997,Petr1998,Simon1999,McCaughrean2001}. The high-mass population exhibits a higher binary frequency than the low-mass primaries \citep{Preibisch1999,Schertl2003,Koehler2006}. For the ONC, we use the study of \citet{Reipurth2007}, who find $75$ binaries and $3$ triples among $781$~ONC members. In the limited range from $67.5$ to $675$~AU, this corresponds to a binary fraction of $8.8\pm1.1$~percent after correcting for contamination. The field binary-fraction for solar-type stars in this range is about $1.5$~times larger. \citet{Tobin2009} add short period binaries to previous studies by using spectroscopic data. They infer a binary fraction of $11.5$~percent in their sample up to a period of roughly $4000$~days, as opposed to $\approx17.5$~percent in the field. Unfortunately, however, they cannot assess their completeness or even report reliable orbital periods.

\subsection{IC 348}
\label{sec:ic348}
IC 348 has a total stellar mass of $110\msun$ \citep{Preibisch2003}, a half-mass radius of $0.47$~pc and an overall stellar density within this radius of $220\mpc$ \citep{Lada1995}. The IMF in IC348 is Salpeter-like above $0.7\msun$ and flattens below that value \citep{Muench2003}, which agrees well with the field-star (i.e. canonical) IMF \citep{Luhman1998,Preibisch2003}. The cluster has a mean age of $2$~Myr, but an age spread of $\approx3$~Myr, and is located at a distance of $320$~pc from the Sun \citep{Herbig1998,Muench2003}.

\citet{Duchene1999b}'s multiplicity data is used in the present work. They investigated the multiplicity properties in a near-infrared survey of $66$~low-mass members of IC 348, finding $12$ binaries in the separation range $\approx40$ to $3200$~AU. The binary fraction of $19\pm5$~percent agrees within the errors with $23\pm3$~percent in the field over the same range of separations.

\subsection{USco-A}
\label{sec:uscoa}
Upper Scorpius (USco) is the youngest ($\approx5$~Myr) region in the Scorpius-Centaurus OB association located at about $145$~pc distance, which contains USco-A. USco-A's stellar IMF between $0.1$~and $20\msun$ is consistent with the field (i.e. canonical) stellar mass function and the total stellar mass is estimated to be about $2060\msun$ \citep{Preibisch2002}. The same authors estimate the total \emph{initial} mass (gas+stars) to be $\approx80000\msun$ and the initial diameter to be $15$~pc given the present-day velocity dispersion.

\citet{Brandner1998} carried out observations of $68$~TTauri stars in USco-A finding $21$~binaries with angular separations of between $15$~and $435$~AU, which is here used for the analysis. \citet{Koehler2000} probe a similar separation range ($19-870$~AU) but have a slightly smaller sample of $48$~TTauri stars. A natural complement to this data is the wide binary survey by \citet{Kraus2009} who identify $21$ binaries between $500-5000$~AU. Here we use the data of \citet{Kraus2008}, who observe $82$~systems finding $27$~companion stars at separations of $6-435$~AU.

\subsection{Praesepe}
\label{sec:praesepe}
The Praesepe open cluster is located $\approx$170~pc from the Sun \citep{Hambly1995} and has an age of about $600$~Myr \citep{Boudreault2010}. \citet{KrausHillenbrand2007} identify $1050\pm30$ Praesepe members earlier than spectral type M5, corresponding to $550\pm40\msun$. The mass-function in Praesepe agrees well with the present-day MF for field stars \citep{KrausHillenbrand2007,Boudreault2010,Baker2010}. \citet{Adams2002} state the half-mass radius to be $1.^{\circ}25$, corresponding to $\approx4$~pc at the distance of Praesepe, and a total mass of $\approx600\msun$ within $4^{\circ}\equiv11.8$~pc.

\citet{Patience2002} find $12$ binaries among $100$ Praesepe members in their sample with separations of $\approx18-390$~AU. The binary fraction is thus $12$~percent, about half as large as in the field. \citet{Bouvier2001}, in contrast, derive a corrected binary frequency of $25.3\pm5.4$~percent, which is in quite good agreement with that measured in the Galactic field. Their work, containing $26$ binaries between $15$ and $600$~AU among their larger sample of $149$ cluster members, is used here.

\subsection{Pleiades}
\label{sec:pleiades}
The Pleiades open cluster is located at a distance of about $125$~pc \citep{Robichon1999} and has an age of about $70-120$~Myr \citep{Stauffer1998,Martin1998}. Its approximately $1245$~member stars have a total mass of about $740\msun$ \citep{ConverseStahler2008,Pinfield1998} and its half-mass radius is $2$~pc \citep{Raboud1998}. Dynamical models of the Pleiades show that the IMF of the Pleiades might have been similar to the Galactic field-star  (i.e. canonical) IMF for stellar masses below $2\msun$ \citep{Moraux2004} and that it may have a precursor similar to the ONC \citep{KroupaAarsethHurley2001}. Its initial binary fraction might have been close to unity \citep{ConverseStahler2010} and its initial distribution of periods similar to that in Taurus \citep{KroupaAarsethHurley2001}.

\citet{Bouvier1997} studied the binary properties of $144$~systems in the Pleiades finding $22$~binaries and $3$~triples in the orbital separation range $10-900$~AU. The corrected binary-fraction of $28\pm4$~per~cent agrees well with the Galactic field value of $27$~per~cent over the same range.\\

\section{Results}
\label{sec:results}
\subsection{The young cluster sample}
\begin{figure*}
 \resizebox{0.9\hsize}{!}{
 $\begin{array}{cc}
   \multicolumn{2}{c}{$\Large${\rm Taurus}} \\
   \includegraphics[width=0.45\textwidth]{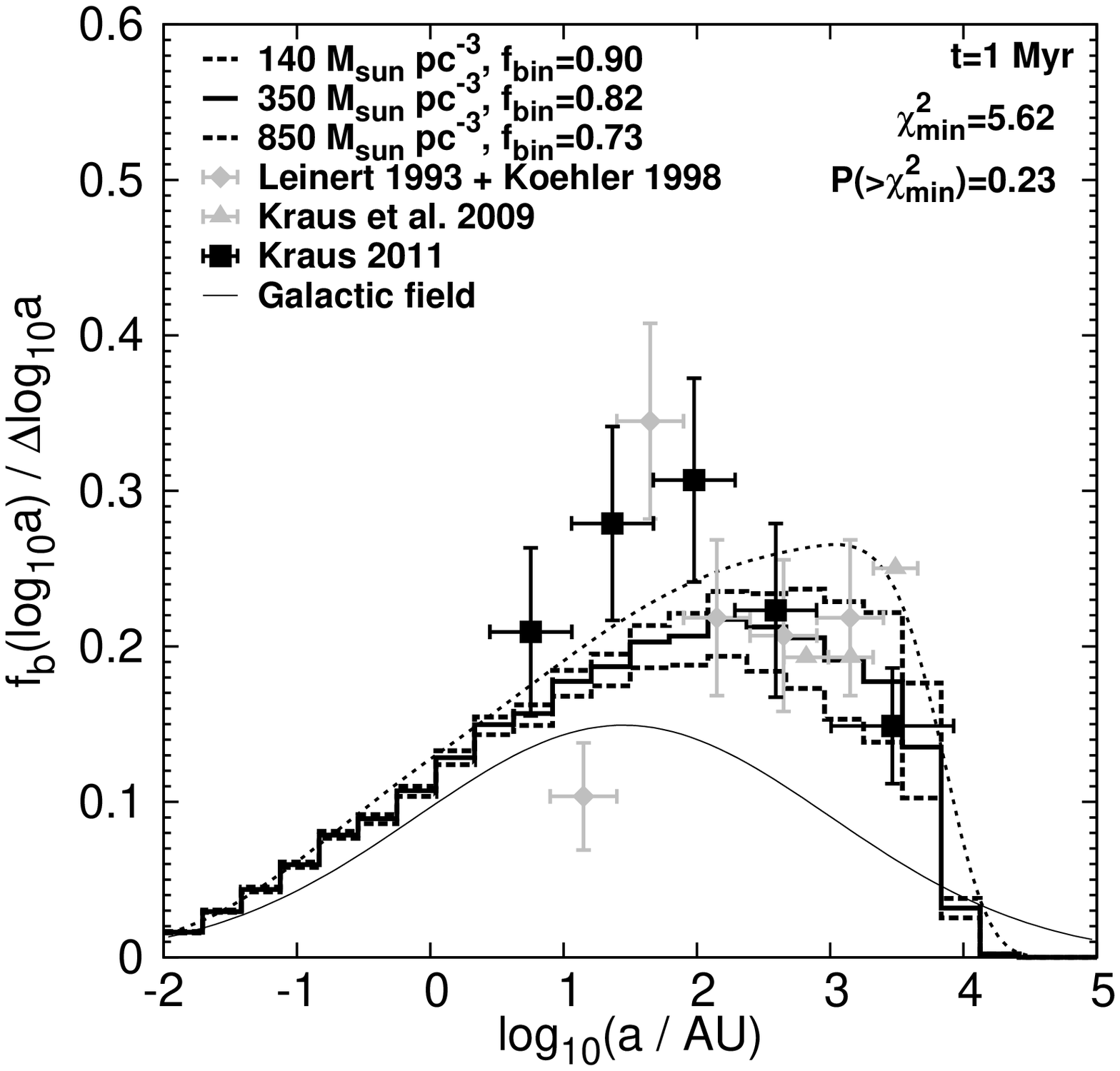} &
   \includegraphics[width=0.45\textwidth]{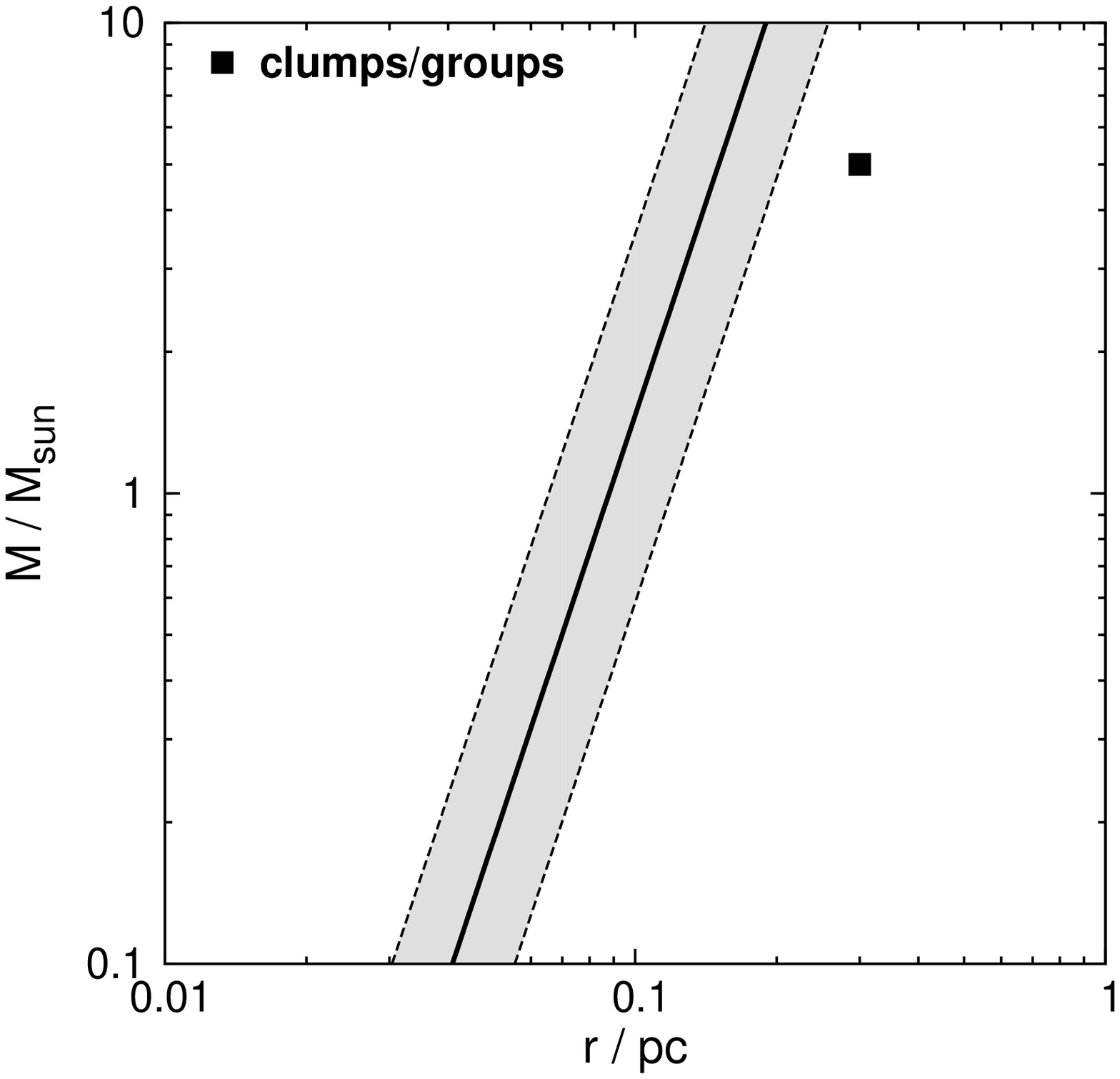} \\
   \multicolumn{2}{c}{$\Large$\rho~{\rm Ophiuchus}} \\
   \includegraphics[width=0.45\textwidth]{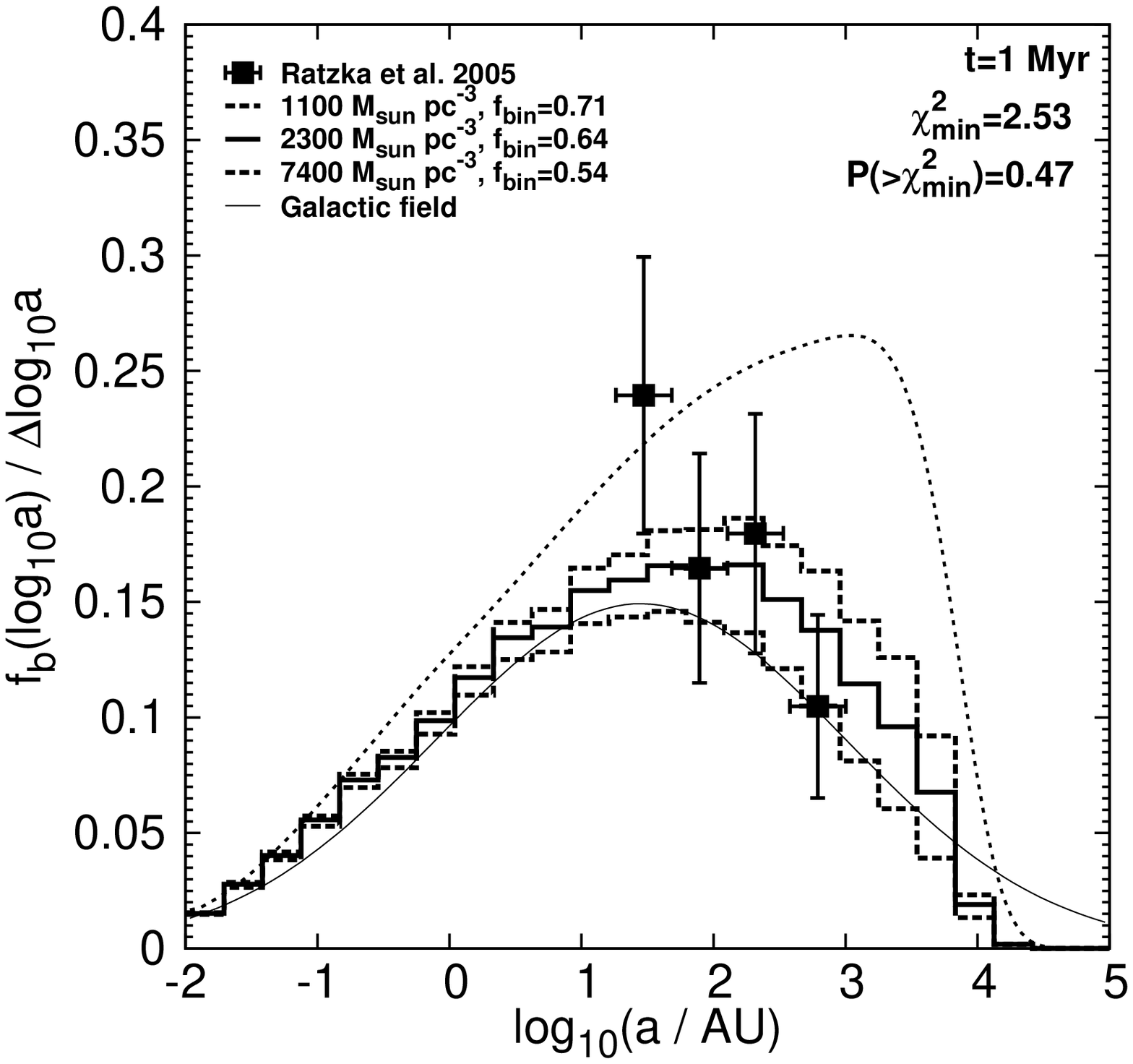} &
   \includegraphics[width=0.45\textwidth]{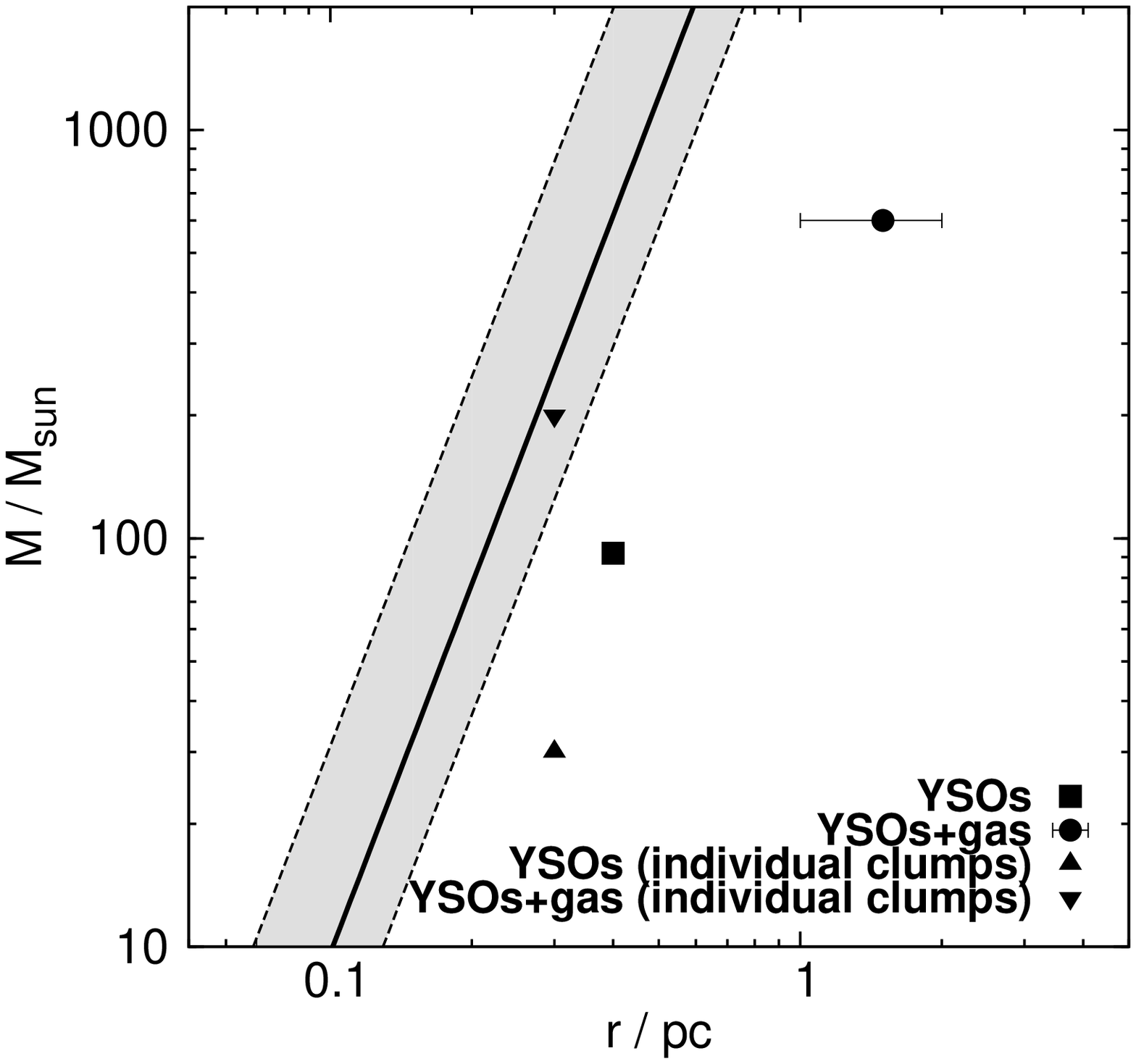}
 \end{array}$
 }
 \caption{\textbf{Left~column:} Semi-major axis distribution functions. Both panels show the initial \citep{Kroupa1995b} semi-major axis distribution in star clusters (dashed curve), as well as the canonical solar-type Galactic field distribution \citep[solid curve,][]{DuqMay1991}. The black squares are observational data from the references given in the legend to which the model is fitted (in Figs.~\ref{fig:taurus_oph}-\ref{fig:praesepe_pleiades} corresponding to the most recent data or the largest sample). Vertical errorbars indicate $\sqrt N$ errors, the horizontal ones indicate the width of the bin. The initial semi-major axis distribution evolves into the solid histogram for an initial stellar density within the half-mass radius, $\rhoecl$, as indicated in the legends. The respective $\chi^2$-tests yield the indicated $\chimin$- and $\probchi$-values, if the initial distribution is evolved for the time $t$ in a cluster of density $\rhoecl$ (note that $t$ is mostly an upper limit to the age of the population). The resulting distributions are likely parent functions for the observational data. The dashed histograms are $10$~percent confidence limits, outside which a distribution would be rejected as a probable parent function with $\gtrsim90$~percent confidence. The overall model binary-fraction, $\fb$ (integrated over the whole semi-major axis range), is indicated in the legend. \textbf{Right~column:} The thick solid and dashed lines are lines of constant volume density within the half-mass radius according to the best-fit $\rhoecl$ and the confidence intervals, respectively. Thus, the model suggests that the total stellar mass and half-mass radius \emph{in a typical, dynamically active region} (e.g. a sub-clump) of the star-forming environment's progenitor lies somewhere in the grey shaded region. For comparison, observed present-day masses and spatial extensions are shown (Sect.~\ref{sec:sample}). Note that the masses and sizes shown denote different spatial scales (e.g. the constant density lines depict the typical total stellar masses, $\mecl$, and the half-mass radius, $\rh$, required to produce the present binary population, while the symbols might refer to different quantities as indicated in the legend - i.e. the total stellar+gas mass within the whole region, not solely individual clumps).}
 \label{fig:taurus_oph}
\end{figure*}
\begin{figure*}
 \resizebox{0.9\hsize}{!}{
 $\begin{array}{cc}
   \multicolumn{2}{c}{$\Large${\rm Chamaeleon}} \\
   \includegraphics[width=0.4\textwidth]{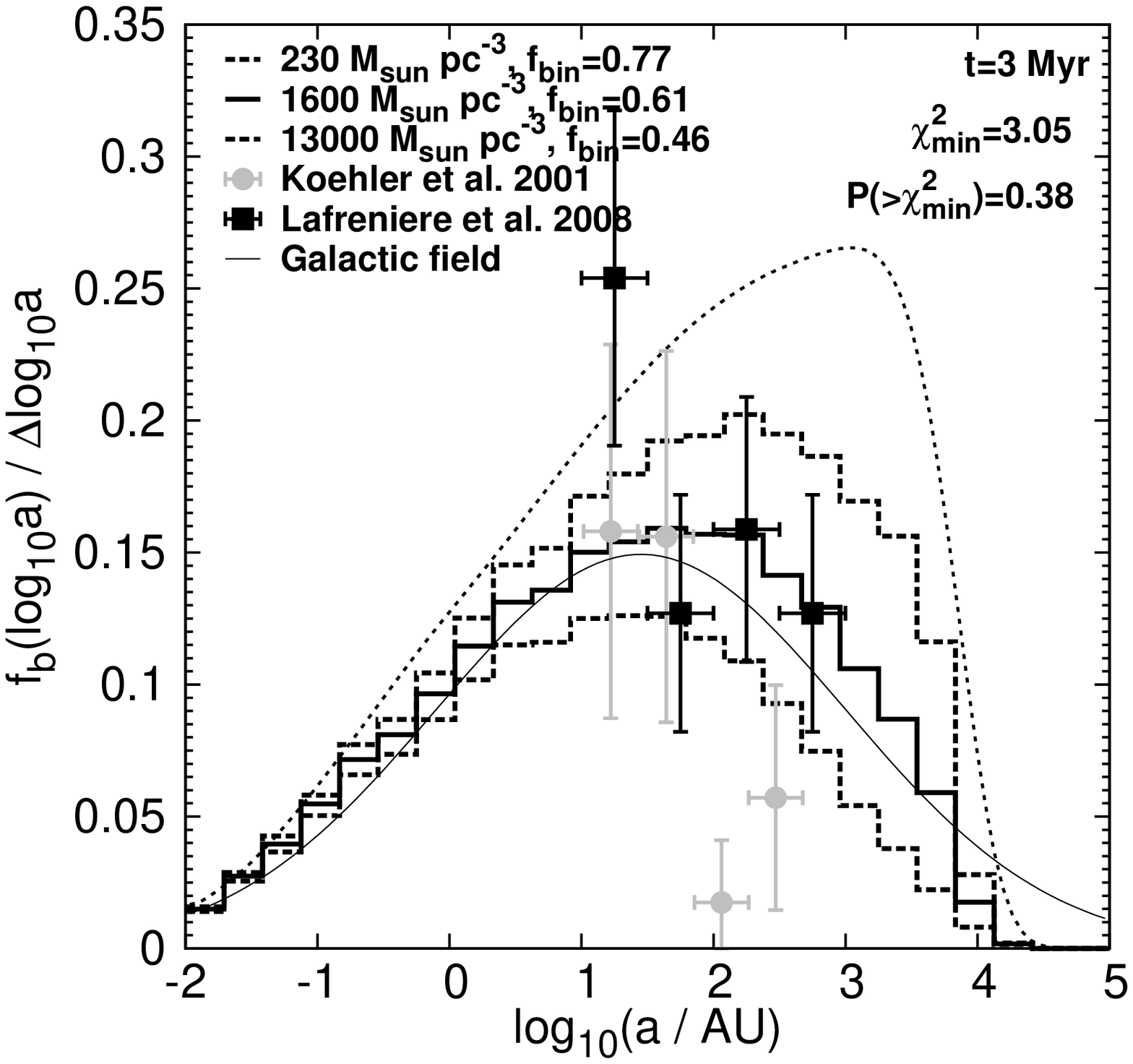} &
   \includegraphics[width=0.4\textwidth]{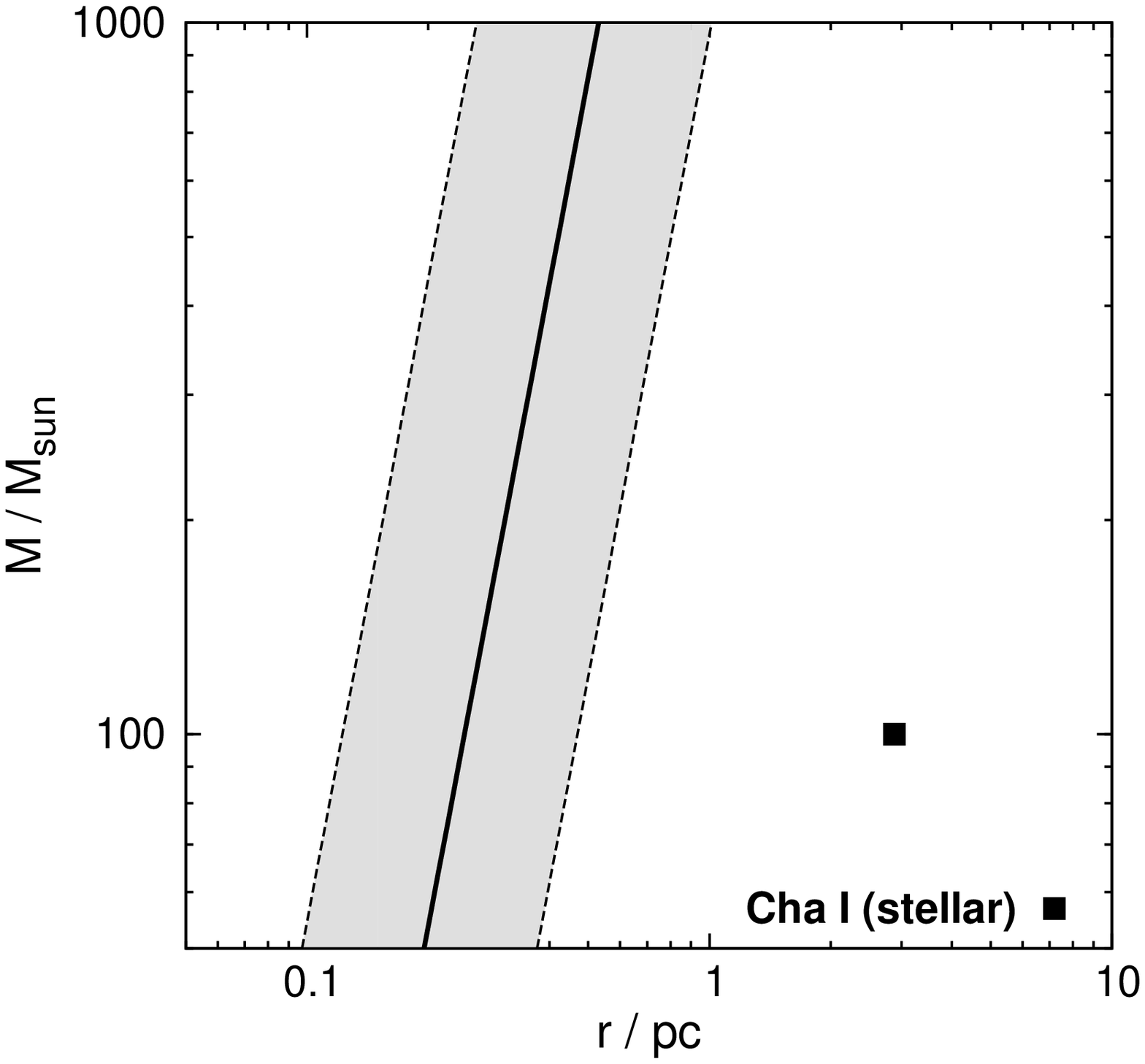}\\
   \multicolumn{2}{c}{$\Large${\rm ONC}} \\
   \includegraphics[width=0.4\textwidth]{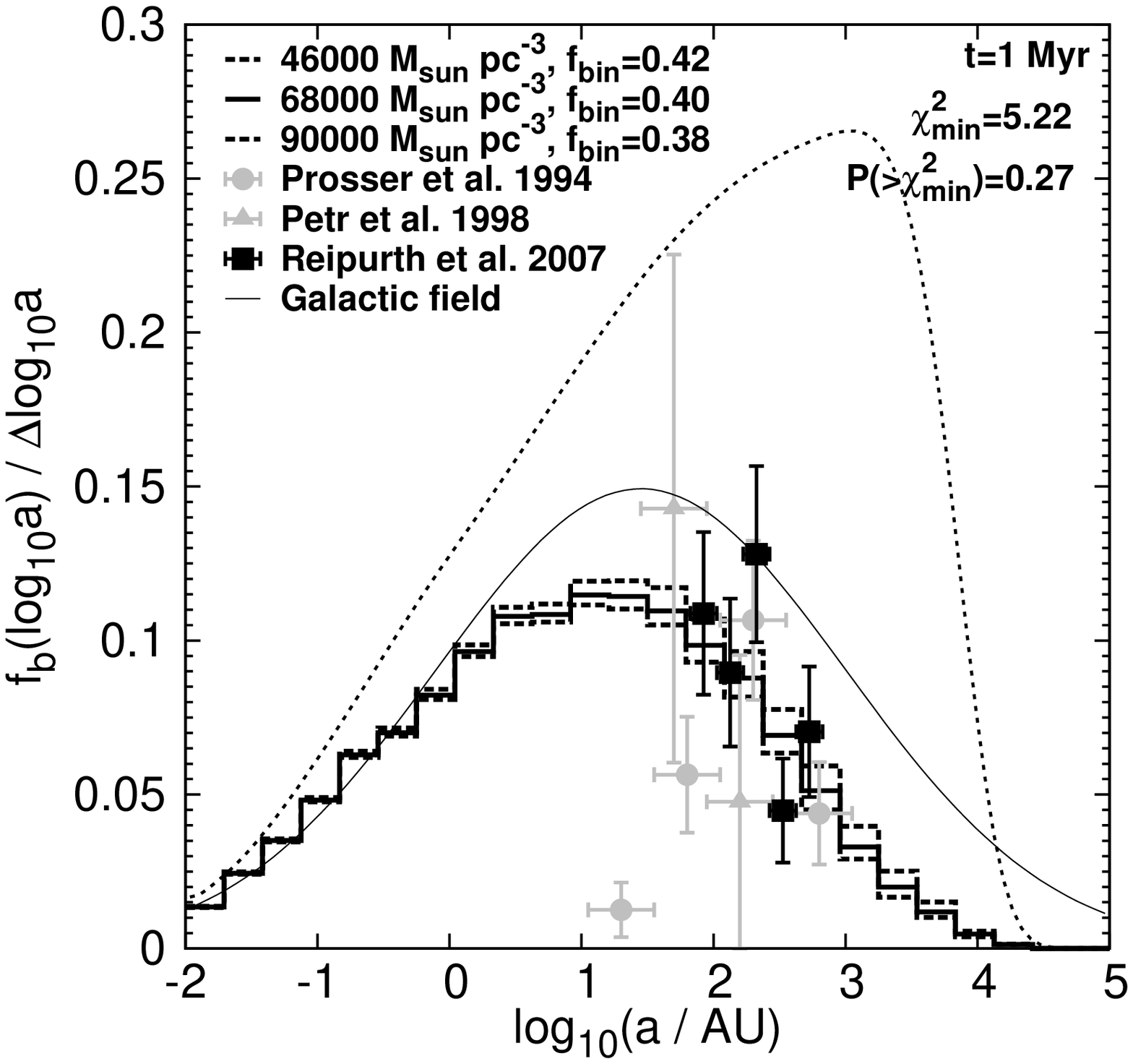} &
   \includegraphics[width=0.4\textwidth]{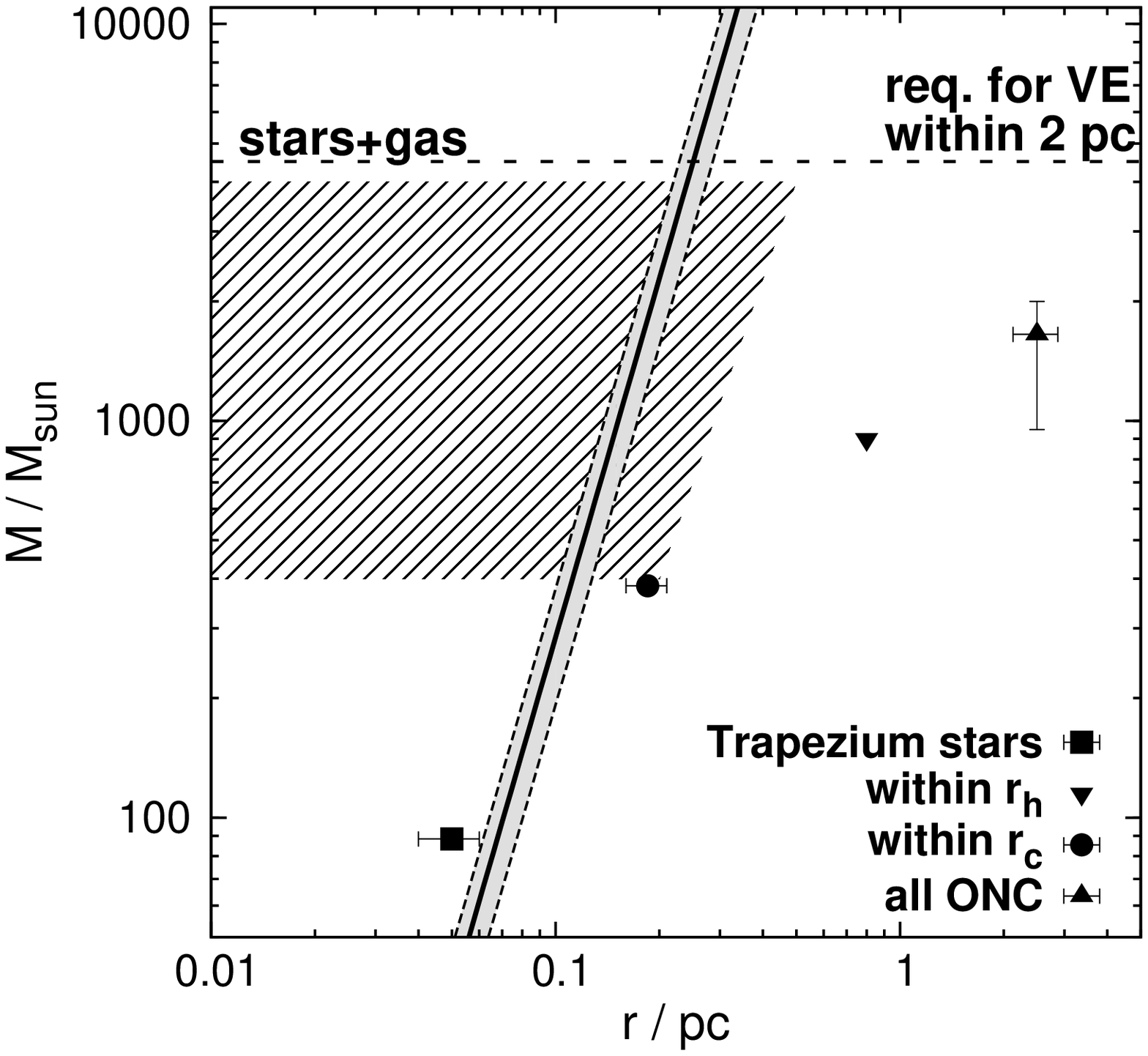} \\
   \multicolumn{2}{c}{$\Large${\rm IC\;348}} \\
   \includegraphics[width=0.4\textwidth]{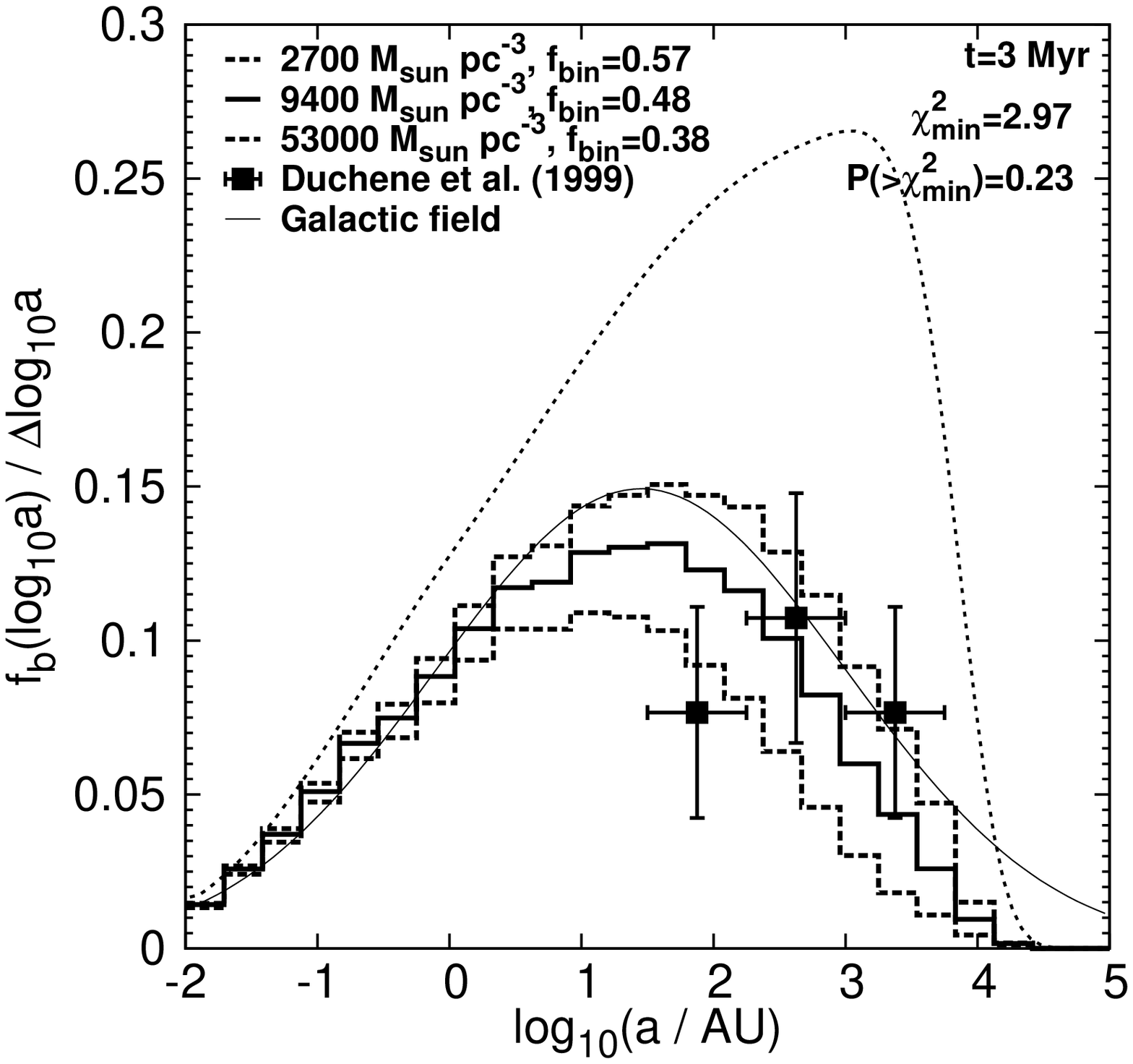} &
   \includegraphics[width=0.4\textwidth]{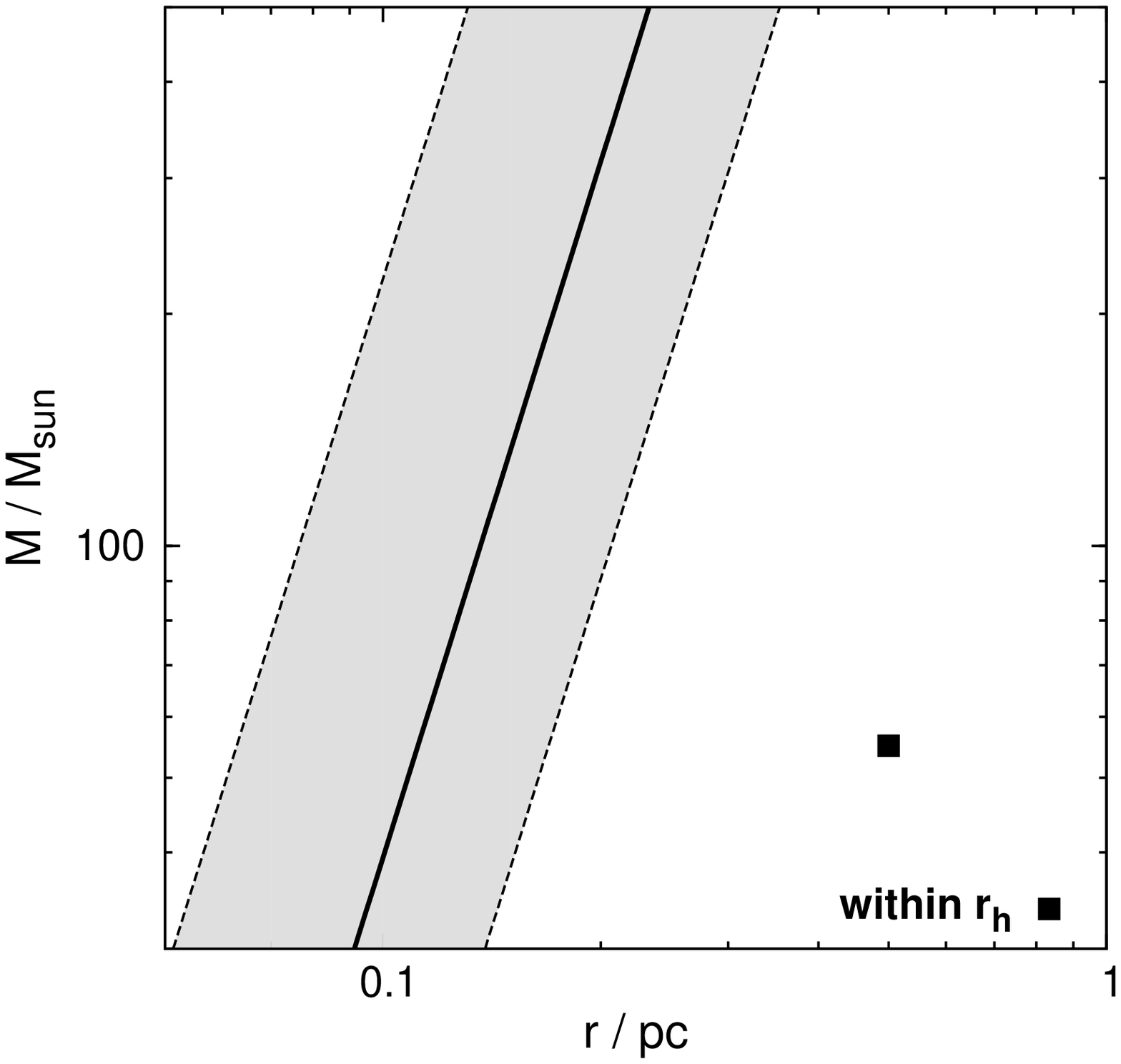}
 \end{array}$
 }
 \caption{As Fig.~\ref{fig:taurus_oph}, but for Chamaeleon, the ONC, and IC 348. The hashed area in the middle-right panel is an allowed region for initial masses and sizes of the ONC according to the expanding cluster models of \citet[his fig.~5]{Kroupa2000}.}
 \label{fig:cham_onc_ic348}
\end{figure*}
\begin{figure*}
 \resizebox{0.9\hsize}{!}{
 $\begin{array}{cc}
   \multicolumn{2}{c}{$\Large${\rm USco A}} \\
   \includegraphics[width=0.4\textwidth]{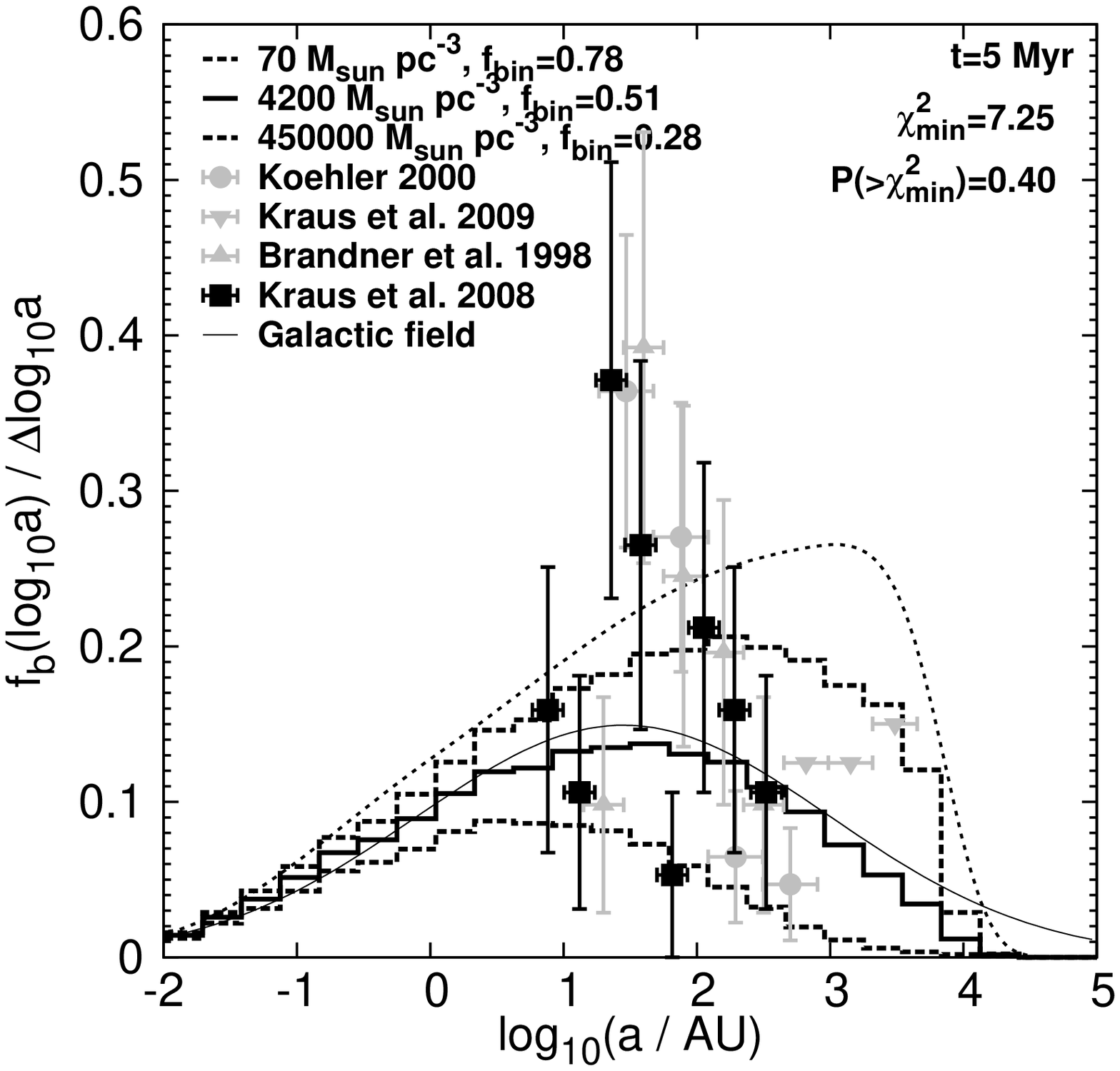} &
   \includegraphics[width=0.4\textwidth]{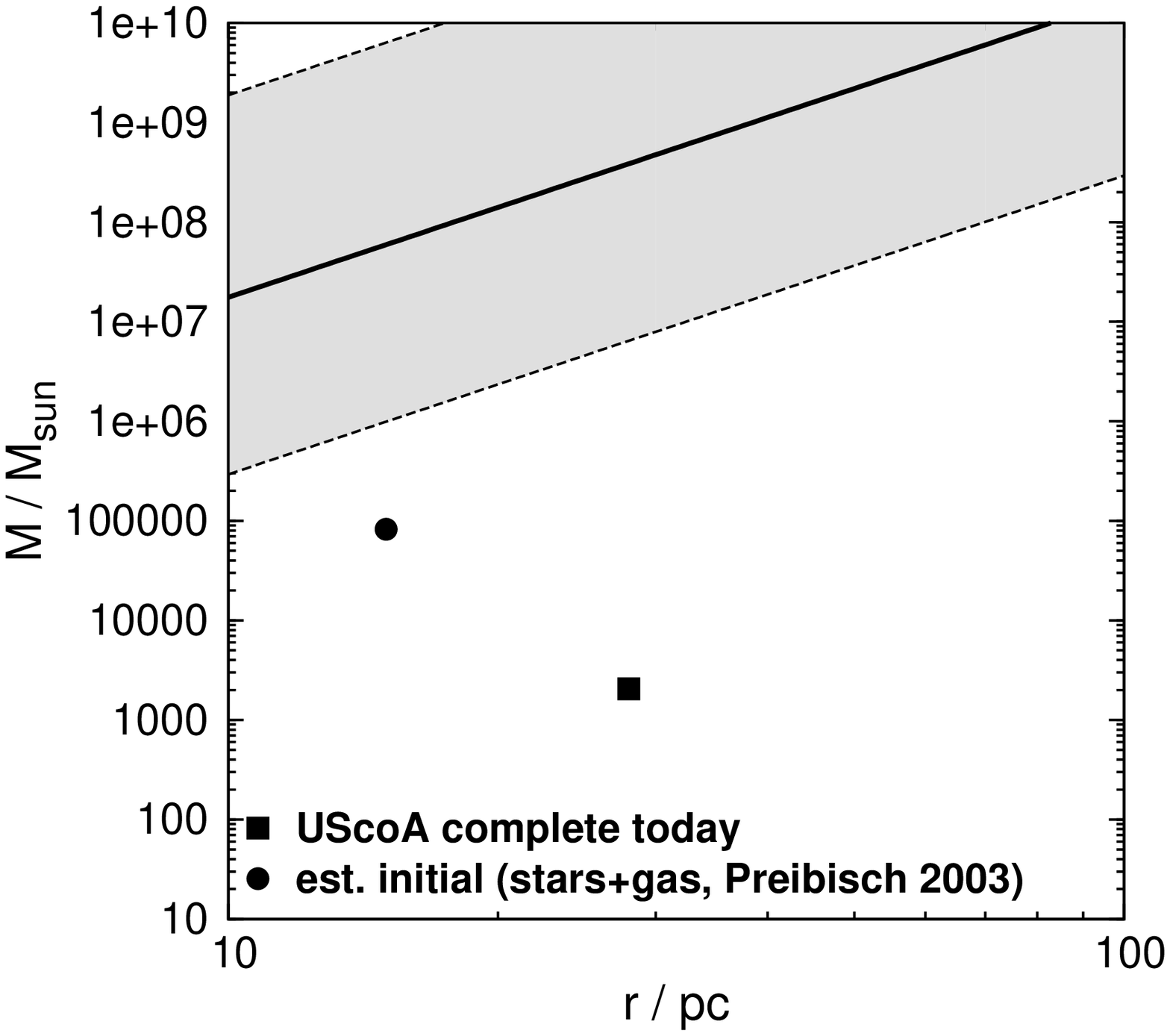}\\   
   \multicolumn{2}{c}{$\Large${\rm Praesepe}} \\
   \includegraphics[width=0.4\textwidth]{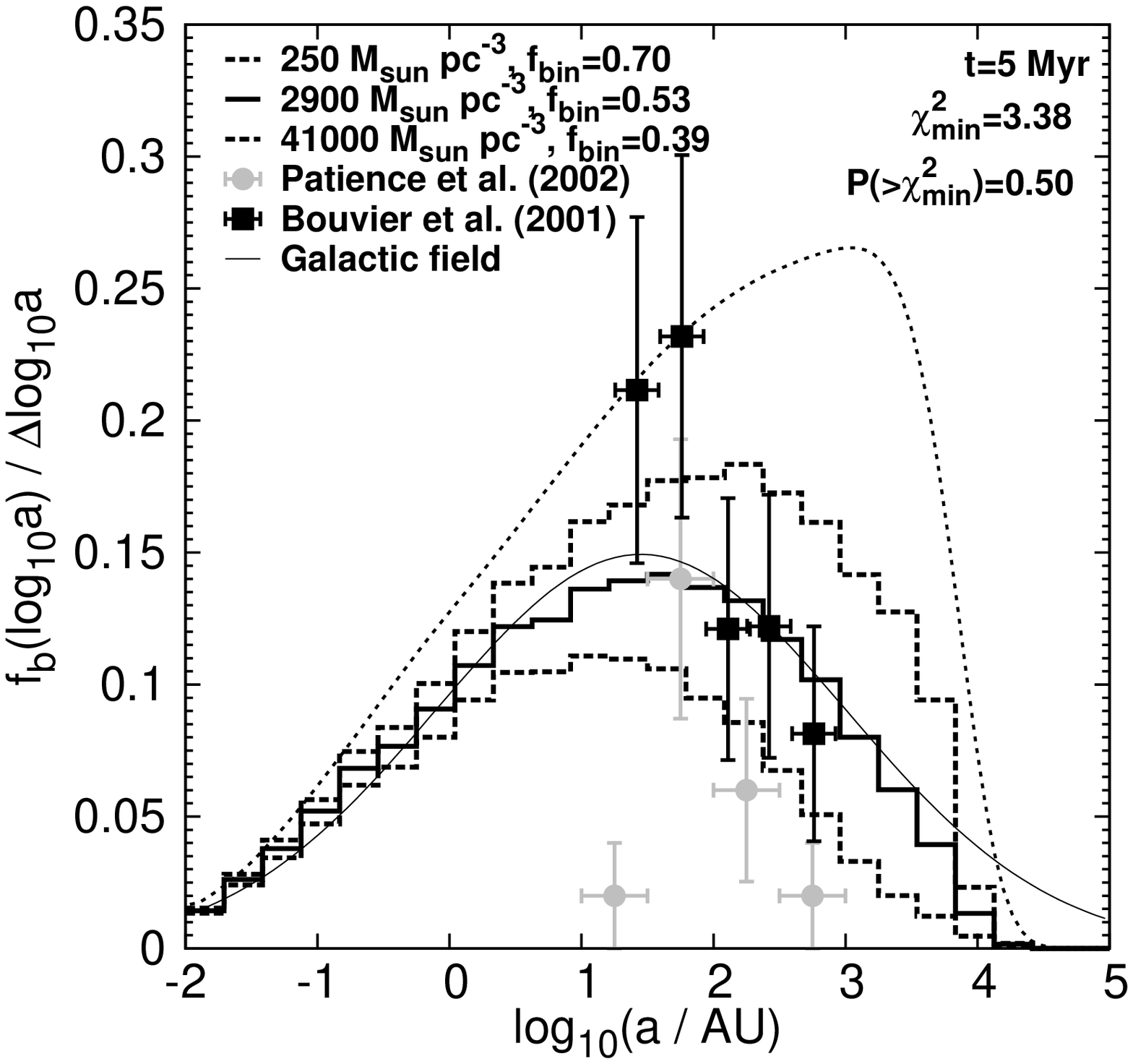} &
   \includegraphics[width=0.4\textwidth]{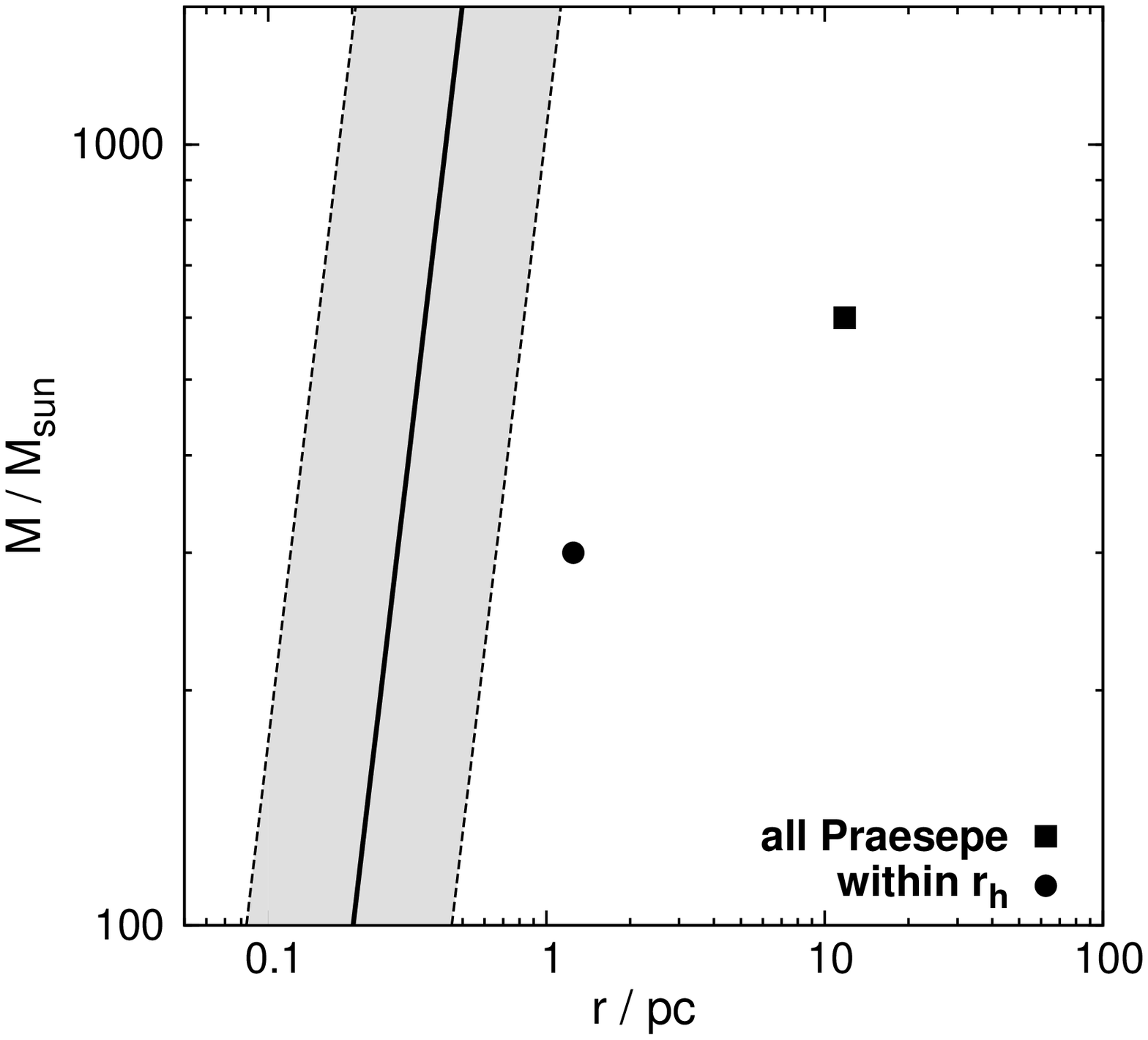}\\
   \multicolumn{2}{c}{$\Large${\rm Pleiades}} \\
   \includegraphics[width=0.4\textwidth]{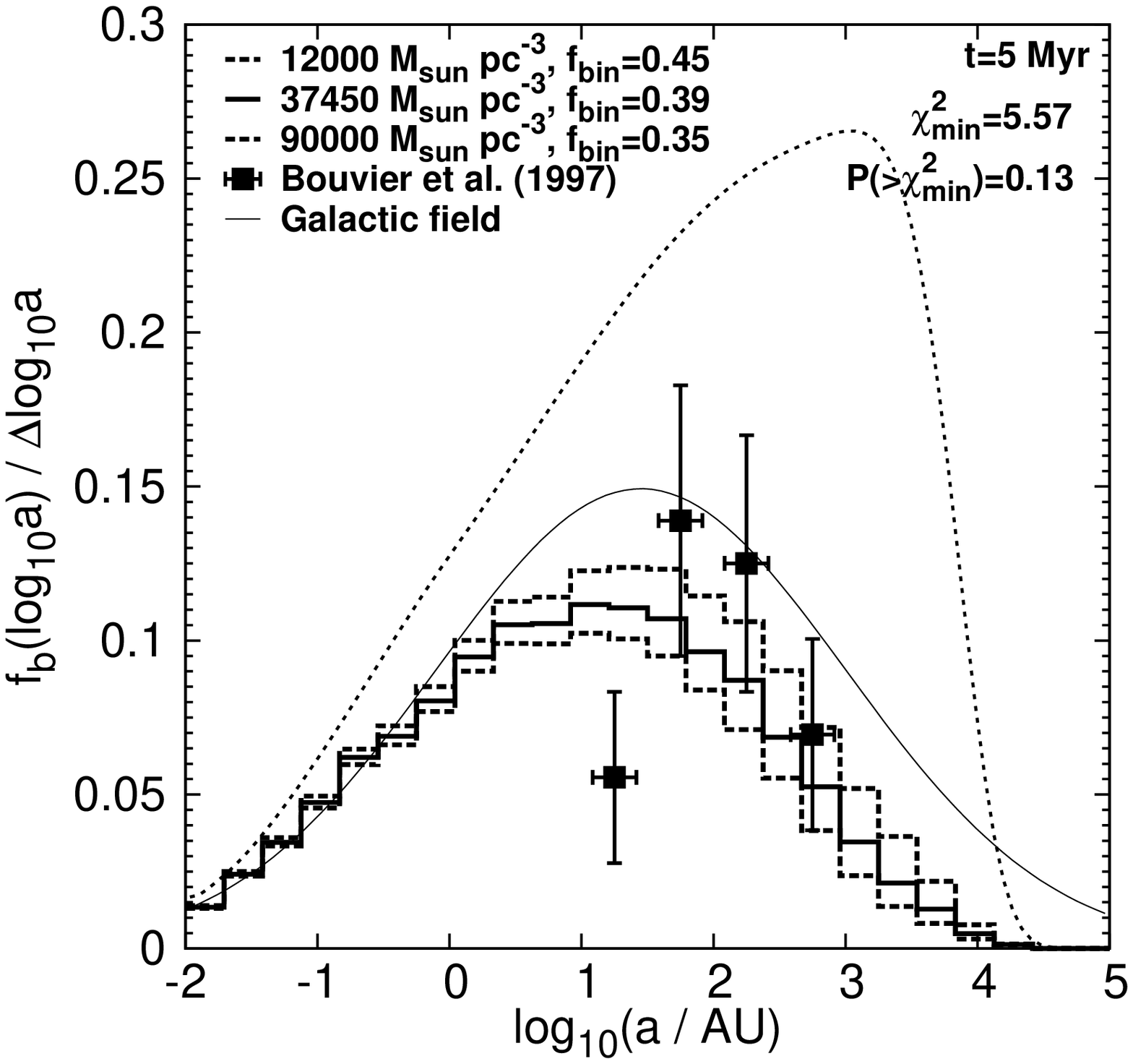} &
   \includegraphics[width=0.4\textwidth]{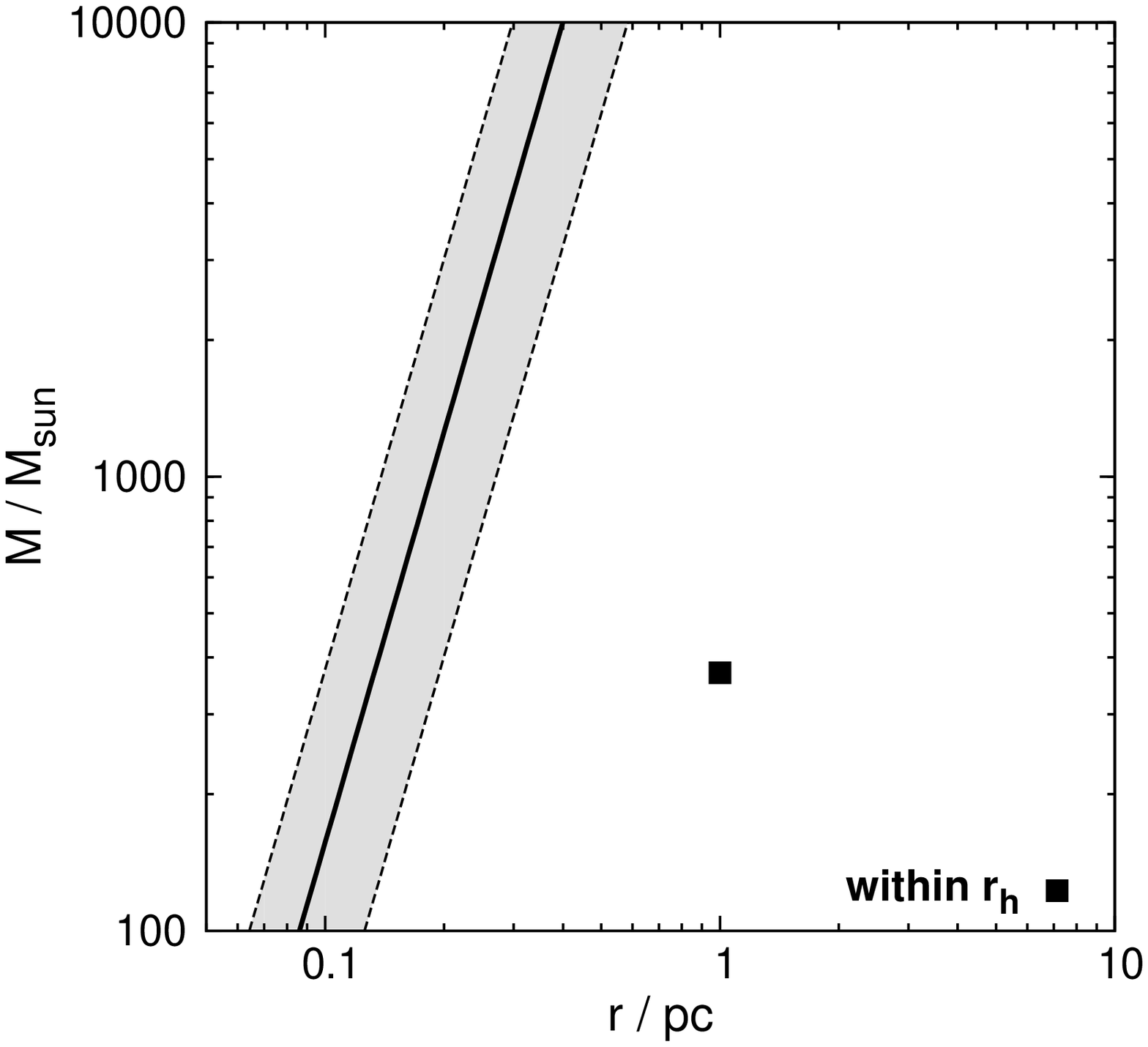}
 \end{array}$
 }
 \caption{As Figs.~\ref{fig:taurus_oph} and~\ref{fig:cham_onc_ic348}, but for USco-A, Praesepe, and the Pleiades. Although Praesepe and the Pleiades are much older ($0.6$~and $0.125$~Gyr, respectively), the initial distribution is evolved for $t=5$~Myr only when binary-burning has largely ended (see the text).}
 \label{fig:praesepe_pleiades}
\end{figure*}
Figs.~\ref{fig:taurus_oph},~\ref{fig:cham_onc_ic348}, and~\ref{fig:praesepe_pleiades} show the results of our analysis. Each best-fit theoretical distribution was evolved for the time indicated in the upper right corner of each panel. This time roughly corresponds to the age of the region (Sect.~\ref{sec:sample}) or if the age is uncertain or shows a spread it is an estimate of the upper age limit.

Although the seven regions exhibit present-day stellar densities ranging from $1-10$~stars~pc$^{-3}$~(Taurus) to $2\times10^4$~stars pc$^{-3}$~(ONC), the model is able to reproduce the observational data for all of them (though only poorly for USco-A, although see below), requiring solely different initial densities. \emph{All dynamically evolved initial binary populations are likely parent functions of the respective observational data according to the $\chi^2$-test.}

It is found that the more depleted the binary population with respect to the initial distribution, the higher the required initial stellar density, $\rhoecl$, to reproduce the observational data (for the same age of the population). Taurus (Fig.~\ref{fig:taurus_oph}, upper panels) is very sparsely populated and still contains a large binary population. Therefore, the initial stellar density was also low, to ensure that the binary disruption was inefficient. According to the model, Taurus sub-clusters formed with $\rhoecl=350\mpc$, but is consistent with a semi-major axis distribution close to the initial distribution within the confidence intervals (Fig.~\ref{fig:taurus_oph}, upper left panel). The current masses and sizes of the individual clusters in Taurus are indeed comparable to the inferred value of $\rhoecl$ (Fig.~\ref{fig:taurus_oph}, upper right panel).

In contrast, the similarly aged $\rho$~Ophiuchus (Fig.~\ref{fig:taurus_oph}, lower panels) and ONC (Fig.~\ref{fig:cham_onc_ic348}, middle panels) show a much stronger depletion of binaries and therefore require higher birth densities for more efficient stimulated evolution. The present-day values are thus farther away from the initial conditions constrained here. This is readily understood as a result of the stronger expansion within the same time of initially denser regions owing to the more effective two-body relaxation and binary burning \citep[fig.~4 in][]{MarksKroupaOh2011} and/or expulsion of gas \citep{KroupaAarsethHurley2001}.

According to our model, the binary fraction above $1000$~AU is two percent for the ONC. For a population of currently approximately $3500$ systems, this would correspond to many more binaries in this semi-major range than the upper limit of three binaries estimated by \citet{Scally1999} for a \citet{DuqMay1991}-type distribution. The model indeed suggests a semi-major axis distribution that is more strongly depleted in binaries (Fig.~\ref{fig:cham_onc_ic348}, middle panels). A detailed study including instrumental limitations and projection effects would need to be performed to address the \citet{Scally1999} analysis. This is however beyond the scope of the present study.

Our model provides only a poor fit to the data for USco~A. Although a fit formally exists, the large confidence intervals would allow a large range of values for the initial density of USco~A. It might thus be difficult to explain this data in terms if dynamical modification of the \citet{Kroupa1995b} initial distribution. USco~A is a complicated region, part of an OB association, and already far more extended than a typical open cluster. It is likely that USco~A may contain projected members of the association. The USco-B subgroup in the Scorpius-Centaurus association is not analysed here, as the binary fraction increases with increasing period \citep{Brandner1998}. This is incompatible with the dynamical evolution of its binary population and may be a contamination effect \citep{KroupaPetr2011}. We note, however, that \citet{Parker2012} caution that binary disruption is a highly stochastic process that influences the appearance of observed binary distributions.

The data for the open clusters Praesepe and the Pleiades (Fig.~\ref{fig:praesepe_pleiades}, $0.6$~and $0.125$~Gyr respectively) are also closely reproduced by the model.\footnote{Note that larger values of $\probchi$ do not signify a statistically tighter fit than for the younger regions, but that one would be less confident that the theoretical distribution is \emph{not} a parent function were the null-hypothesis to be rejected.} The initial distribution for the best-fit $\rhoecl$ was evolved for only $5$~Myr, but by then the evolution of the binary distributions had effectively come to an end \citep{MarksKroupaOh2011} so that a comparison with the much older regions should be possible. That the binary population does not evolve significantly after 3-5~Myr is clearly evident from a comparison of \citet{MarksKroupaOh2011} with \citet{KroupaAarsethHurley2001}.

\begin{figure}
 \resizebox{\hsize}{!}{
   \includegraphics[width=0.45\textwidth]{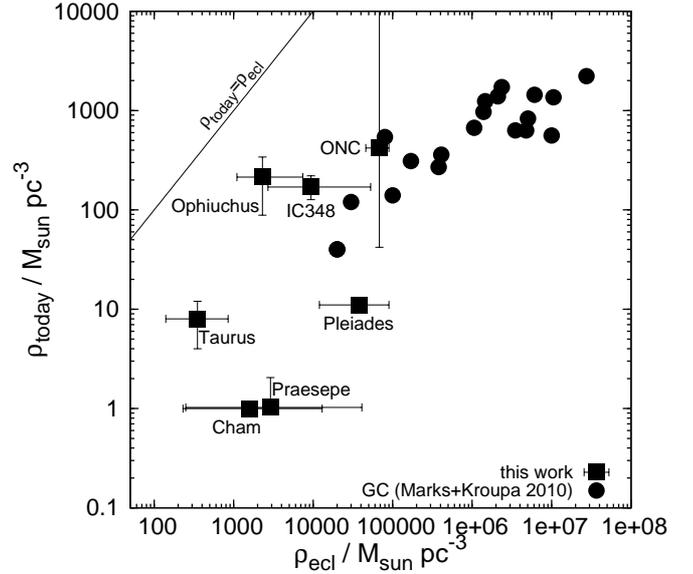}
 }
 \caption{Comparison of present-day and inferred initial stellar densities of \emph{typical regions}. All seven regions evolve to lower densities away from their initial values. Ranges for the present-day densities are calculated from the total stellar masses and half-mass radii reported in Sect.~\ref{sec:sample}. USco-A is not plotted here as no present-day radius with a given enclosed mass is reported in the literature that could reasonably be compared to the half-mass densities of its initial configuration constrained here. Note that some mass estimates only cover a limited spectral-type range. The filled circles indicate the deduced initial conditions for some Galactic GCs \citep{MarksKroupa2010}.}
 \label{fig:densities}
\end{figure}
According to the model, for IC 348 and the Pleiades the best-fit initial densities are very similar, as are those for Praesepe, $\rho$~Ophiuchus, and Chamaeleon (Fig.~\ref{fig:densities}, not considering the rather large confidence intervals). Taurus and the ONC occupy the low- and high-density end, respectively. The large confidence intervals even allow Praesepe and the Pleiades to have formed with the ONC density, as discovered before by \citet{KroupaAarsethHurley2001} using independent arguments.

\subsection{Cluster evolution depending on density and comparison with the initial conditions of globular clusters}
\citet{MarksKroupa2010} provide independently obtained constraints of the initial conditions for some Galactic globular clusters (GCs). They use the imprint of residual-gas expulsion still visible in the present-day stellar mass function of GCs to backtrace the present cluster properties to their initial values. Their results neatly join the densest young clusters investigated in Fig.~\ref{fig:densities}, suggesting a connection between the initial and present-day densities of clusters, although this is difficult to infer from the young cluster sample alone. \citet{MarksKroupaOh2011} indeed find that a dense initial configuration can be less dense \emph{today} than another initially less dense cluster.

\begin{figure*}
 \resizebox{\hsize}{!}{
   \includegraphics[angle=-90,width=\textwidth]{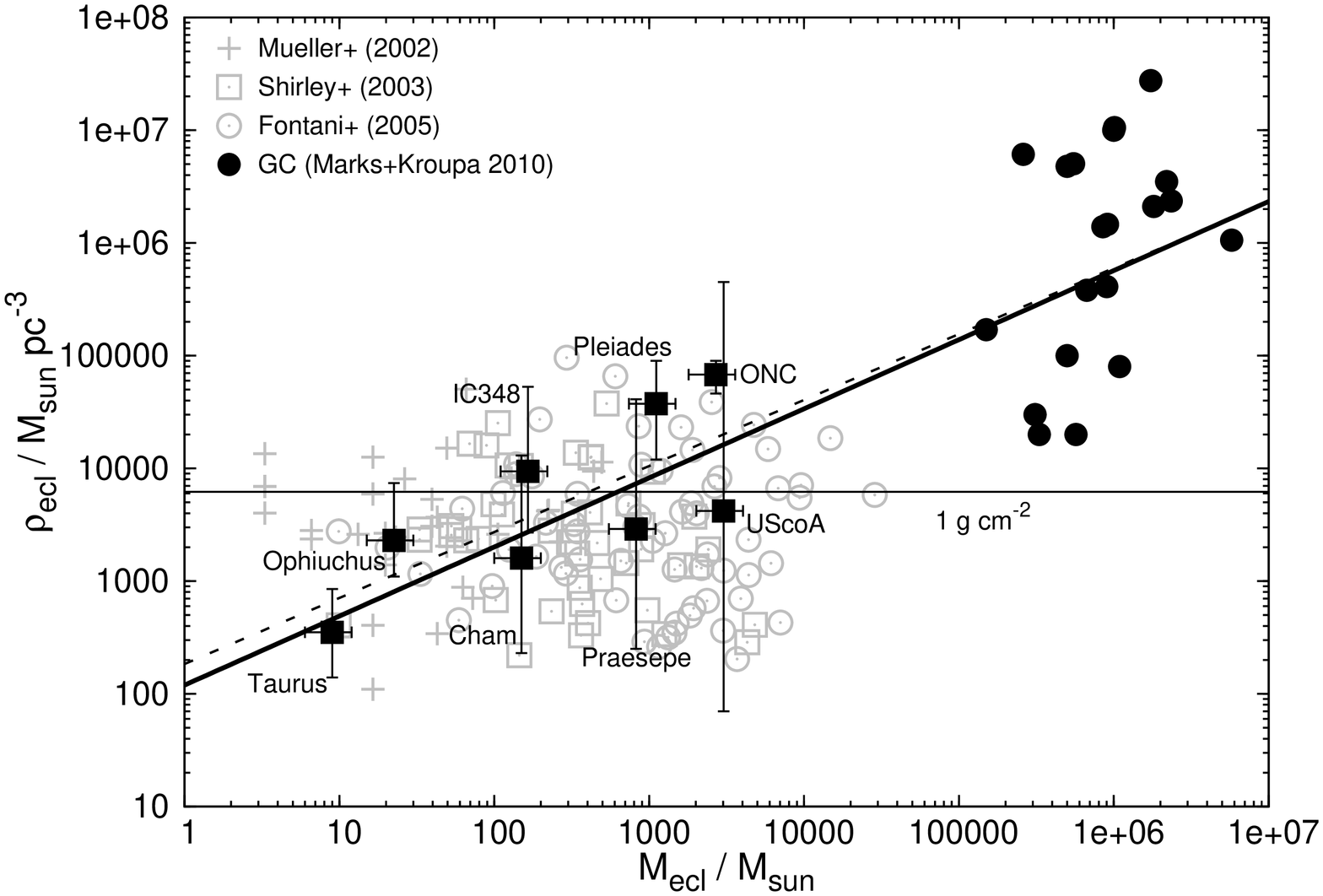}
 }
 \caption{Constraints on the initial volume-densities within the half-mass radius and masses derived in this work for the seven clusters (filled squares) versus the initial stellar mass. The indicated errors in mass correspond to the observationally inferred present-day mass on the left end of a bar and two times the present-day mass on its right end, to be understood as an estimator of the possible initial-mass range. Filled circles are Galactic GCs as in Fig.~\ref{fig:densities}. Underlaid as grey symbols are data of molecular cloud clumps of \citet[crosses]{Mueller2002}, \citet[squares]{Shirley2003}, and \citet[circles]{Fontani2005} as collated by \citet{Parmentier2011}. These are known to have already begun forming stars. The clump masses have been multiplied with a star-formation efficiency of one-third to compare to the stellar masses and densities inferred in the present work. The thin solid black-line is the threshold for massive-star formation evaluated by \citet[$1$~g~cm$^{-2}$, see the text]{Krumholz2008}. The thick solid black-line is a least squares fit to both the young cluster and GC data (eq.~\ref{eq:fit}), implying that there is a mass-radius relation (eq.~\ref{eq:massradius}) for star cluster-forming cloud clumps. The dashed line shows the result when the GCs are excluded from the fit.}
 \label{fig:clumps}
\end{figure*}
Fig.~\ref{fig:clumps} shows that the constraints compare well with masses and densities of molecular clumps that have just begun to form stars \citet{Mueller2002,Shirley2003,Fontani2005}. The masses of the clumps from these sources have been multiplied with a star-formation efficiency of one-third for comparison with the stellar masses and densities calculated here.

Except for Taurus, all of the investigated regions lie close to or above the threshold of the surface density of massive-star formation of $1$~g~cm$^{-2}$ suggested by \citet[which is the same for all cluster masses]{Krumholz2008}. This value corresponds to $4788\msun$~pc$^{-2}$ and, assuming an average pre-cluster molecular cloud diameter of $0.76$~pc, the average clump size in the three aforementioned observational data sets, corresponds to $\rhoecl=4788/0.76=6202\mpc$ (thin solid line, for a homogeneous mass distribution within the clumps).

According to this interpretation, far more-massive and denser clumps were needed to form proto-GCs. Considering our derived constraints (solid squares in Fig.~\ref{fig:clumps}) and the data sets for GCs (solid circles) we find a relation between stellar mass-density and stellar mass of the form
\begin{equation}
 \log_{10}\rhoecl=a\times\log_{10}\mecl+b\;,
 \label{eq:fit}
\end{equation}
where $a=0.61\pm0.13$ and $b=2.08\pm0.69$ (thick solid line). Fitting only to the young cluster data $a=0.59\pm0.22$ and $b=2.26\pm0.57$ (thick dashed line), i.e. both fits are indistinguishable. We note however that the trend without the GCs is mainly driven by Taurus and the ONC. Excluding these two only a rather flat initial stellar mass-density--stellar mass relation emerges. Equation~\ref{eq:fit} implies that there is a weak stellar mass--half-mass radius dependence of star clusters at birth of the form
\begin{equation}
 \frac{\rh}{{\rm pc}}=\sqrt[3]{\frac{3(\mecl/\msun)^{1-a}}{8\pi\cdot10^b}}=0.10_{-0.04}^{+0.07}\times\left(\frac{\mecl}{\msun}\right)^{0.13\pm0.04}\;,
 \label{eq:massradius}
\end{equation}
which is consistent with a relation between the effective radius and star cluster mass for a sample of clusters in non-interacting spiral galaxies, $R_{\rm eff}\propto M^{0.10\pm0.03}$, found by \citet{Larsen2004}. \citet{Zepf1999} and \citet{Scheepmaker2007} similarly demonstrated that there is a mass -- luminosity relation for clusters that is shallow for both the galaxy merger NGC~3256 and young clusters in M51.

\begin{figure*}
 \resizebox{\hsize}{!}{
   \includegraphics[angle=-90,width=\textwidth]{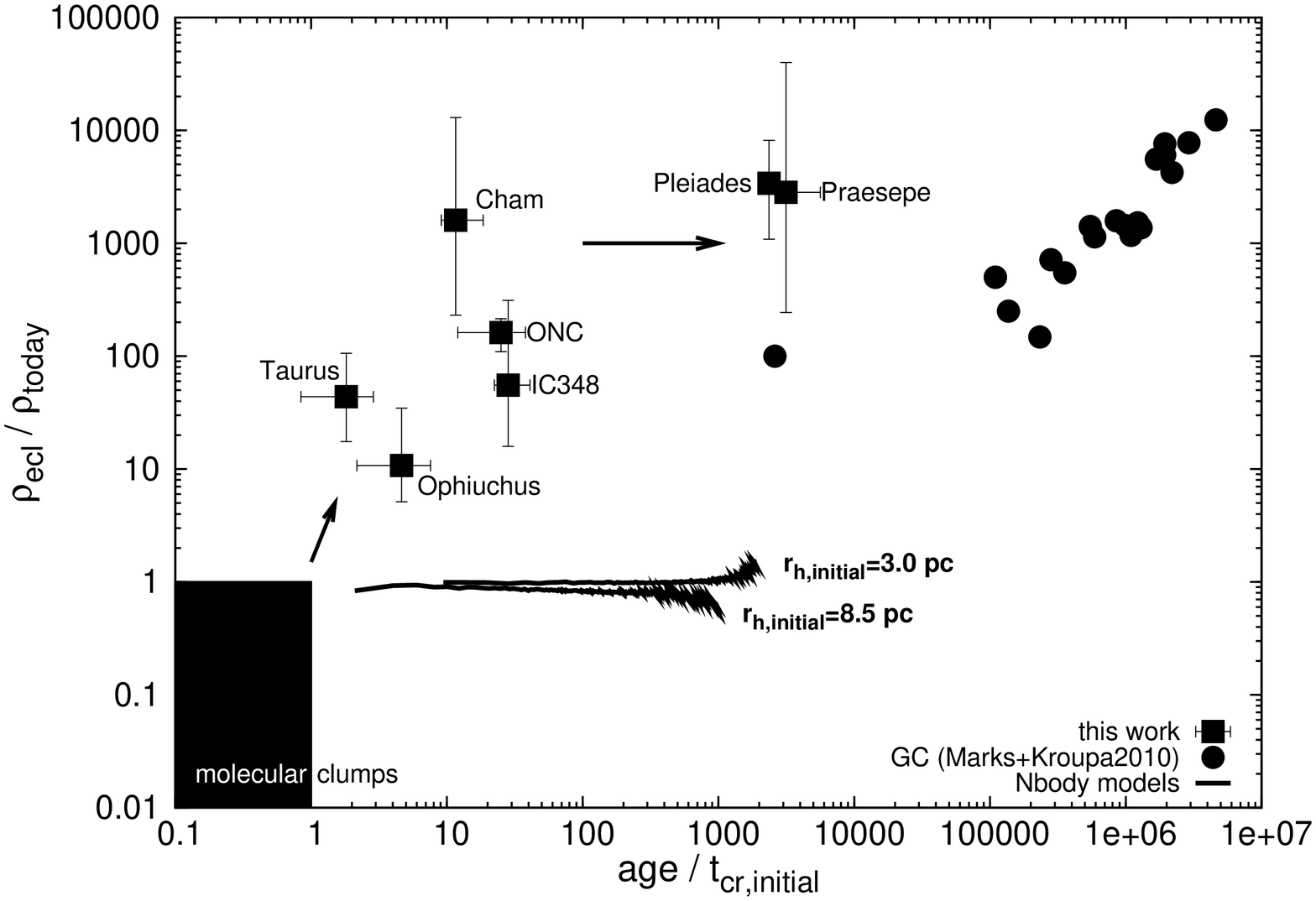}
 }
 \caption{The ratio of the constrained initial to the present-day density as a function of the number of completed initial crossing-times, $\tcri$, for seven clusters (filled squares, where USco-A is not shown for the same reason as it is not depicted in Fig.~\ref{fig:densities}). Errors in the x- and y-directions follow from the error in $\rhoecl$ and error propagation. These clusters may originate from the bottom-left corner of the diagram, the region supposedly occupied by the molecular clumps shown in Fig.~\ref{fig:clumps}. The clumps are probably still collapsing and have not yet reached their peak densities, hence have a density ratio $<1$. Filled circles indicate the constraints for Galactic GCs as in Figs.~\ref{fig:densities} and ~\ref{fig:clumps} \citep[ages taken from][]{MarinFranch2009}. The solid arrows indicate a possible evolutionary sequence. This evolution cannot be understood in terms of single-star $N$-body star cluster models without gas expulsion. These models are shown by the evolution of the $\mecl=2\times10^4\msun$ clusters with the indicated initial-$\rh$ and a \citet{Kroupa2001} IMF between~$0.1\msun$ and~$1.2\msun$ on circular orbits around a realistic Galactic potential over $4$~Gyr \citep[solid black curves]{Kuepper2010}. Residual-gas expulsion and/or energy generation through binary-burning probably needs to be invoked to explain the expansion of the real clusters (see the text).}
 \label{fig:evolution}
\end{figure*}
In Fig.~\ref{fig:evolution}, the clusters that we have studied appear to follow an evolutionary sequence that depends on the number of completed initial crossing-times,
\begin{equation}
 \tcri=\frac{2}{\sqrt{G}}\;\mecl^{-1/2}\;\rh^{3/2}\;.
\end{equation}
The Galactic GCs \citep{MarksKroupa2010} are older and, thus, dynamically more evolved but show similar initial-to-present-day density ratios as Chamaeleon, Praesepe, and the Pleiades. The molecular clumps presented in Fig.~\ref{fig:clumps} are expected to lie at the bottom-left corner of Fig.~\ref{fig:evolution} as these continue to collapse before star formation sets in.

The results presented here suggest that bound star clusters reduce their birth density by a factor of a few hundred within $10-100$~initial crossing times and then evolve at about constant density. The standard secular evolution tracks (without gas and binaries) as predicted by $N$-body models \citep{Kuepper2010} with initial conditions as stated in the caption of Fig.~\ref{fig:evolution} are unable to explain the constrained location for the young clusters and GCs. The energy that initially expands clusters considerably instead needs to be produced to achieve consistency with the young cluster and GC data presented here. This production may be possible by residual-gas expulsion and/or the burning of binaries, which enables the freshly-hatched clusters to ascend the track described by the real clusters in this work. The most massive clusters may evolve into old GCs, but, after their gas expulsion- and binary-burning-driven expansion, at a slower pace in terms of the evolution of their density, which is probably similar to the quiescent evolution of the presented $N$-body models. In this sense, GCs and open clusters (Pleiades, Praesepe) may have formed via a single cluster formation mode, as previously suggested by \citet{Larsen2002}. The intrinsic slope evident for the GCs in Fig.~\ref{fig:evolution} may however suggest that the picture is more complicated. This needs to be tested by self-consistent $N$-body models of star clusters including gas expulsion and binaries as cluster heating sources.

\section{Discussion}
\label{sec:sum}
We have shown and discussed that evolving an assumed universal initial binary population in star clusters for the age of the considered object leads to theoretical semi-major axis distributions that are consistent with the observed ones in six young clusters and star-formation regions, as well as in two older open clusters. Only the stellar densities at birth we needed to carefully select to adjust the strength of stimulated evolution. For all investigated objects, the theoretical semi-major axis distributions turn out to be parent functions of observational data.

\citet{MarksKroupaOh2011} demonstrated that there is an \emph{age-density degeneracy}, i.e. regions of different age can have the same binary population if their birth densities are properly adjusted. This degeneracy is lifted here because the investigated populations have a reasonably well-known age allowing IDPS to to be applied to each cluster so that the initial density of a typical regions' progenitor (i.e. a $\approx0.5$~pc sub-clump or individual cluster in the case of Taurus but the whole ONC) can be inferred. Taurus appears to host a largely unevolved binary population, in agreement with explicit $N$-body modelling \citep{Kroupa2003}. The present-day observed masses and sizes of individual $\approx0.5$~pc clumps in Taurus are therefore close to what is expected based on IDPS of the regions' initial conditions. In the case of the ONC we need to assume the highest density at birth to turn the initial separation distribution into the observed one and its present-day mass and radius are rather far away from the ones at birth. In-between these two extremes, IC~348 and the Pleiades on the one hand and Praesepe, $\rho$~Ophiuchus, and Chamaeleon on the other share very similar initial densities according to this analysis. For each cluster, the calculated initial density agrees with the pre-stellar cloud clump densities observed in molecular clouds (Fig.~\ref{fig:clumps}). This therefore suggests that the expansion caused both by residual-gas expulsion and efficient energy generation owing to binary-burning increases with the ratio of age to $\tcri$ (Fig.~\ref{fig:evolution}).

Furthermore, by adding the constraints on the birth densities for Galactic GCs obtained by \citet{MarksKroupa2010}, we uncover a stellar mass -- half-mass radius relation of star-forming cloud cores (eq.~\ref{eq:massradius}). This relation implies that the radii of birth clusters have a weak stellar mass dependence, $\rh\propto\mecl^{0.13\pm0.04}$, that is consistent with other observational data \citep{Zepf1999,Larsen2004,Scheepmaker2007}.

\citet{King2012} compare the binary properties of five young clusters analysed here (except USco-A). They find no correlation between the present-day binary fraction and the \emph{present-day} (instead of initial) number density, at least if Taurus is not considered. The present-day density might however not necessarily reflect the initial conditions as the large density of an initial configuration (with strong binary burning) can drop below the density of an initially less dense configuration within only a few Myr \citep[cf.~fig.~4 in][]{MarksKroupaOh2011}. Residual-gas expulsion may enhance this effect.

Using $N$-body computations, \citet{King2012} tentatively argue for fractal initial conditions in these regions, even though this is not the best $N$-body model for all their studied regions. They invoke this substructure on different scales ($\lesssim1$~pc for ONC- and IC348-type clusters, but $\lesssim10$~pc for Taurus-type configurations). The sizes of the clumps in Taurus are however comparable to the extent of the ONC and IC348, as evident in their figs.~7-11. Thus, there appears to be \emph{one dynamically relevant scale}, i.e. the scale on which dynamical (binary) evolution is important, of about $\approx0.5-1$~pc. This size corresponds well to the sizes in the $N$-body models on which the \citet{MarksKroupaOh2011} analytical model is based and that are used here to derive the initial densities.

The analysis here finally suggests that stimulated evolution plays an important role in shaping the present-day properties of binary populations in star clusters. There exists of a formal initial binary distribution function that is consistent with the available multiplicity data of pre-main sequence and Class I protostellar populations \citep{KroupaPetr2011} and that helps us to understand the binary distributions in the eight objects investigated here simultaneously, assuming dynamics only. This suggests that there are no significant variations in binary-star formation depending on the ambient condition required, but that variations in the binary populations are mainly caused by stellar and dynamical evolution. This agrees with previous work and such an assumption would be similar to the universality hypothesis of the stellar IMF \citep{Kroupa2011,KroupaPetr2011}.

The alternative is that the birth binary populations depends on the cloud density such that a larger fraction of binaries form in low-density environments. This is a trivial solution to the observed trend. However, the IMF is known to be invariant for the star-forming conditions probed here \citep{Kroupa2012}, and it is unclear how the IMF of systems could be separated from the IMF of all individual stars if there were a variation in the properties of the initial binary population.

\begin{acknowledgements}
MM was supported for this research through a stipend from the International Max Planck Research School (IMPRS) for Astronomy and Astrophysics at the Universities of Bonn and Cologne. We thank Genevieve Parmentier for providing the molecular clump data in Fig.~\ref{fig:clumps} in machine-readable format, as compiled in \citet{Parmentier2011}.
\end{acknowledgements}

\bibliographystyle{aa}
\bibliography{binaries}

\end{document}